\documentclass[
 aps,
 pra,
 reprint,
 superscriptaddress,
 nofootinbib,
 amsmath,
 amssymb
]{revtex4-2}

\usepackage[T1]{fontenc}
\usepackage{amsthm}
\usepackage{mathtools}
\usepackage{stmaryrd}
\usepackage{mathrsfs}
\usepackage{graphicx}
\usepackage{dcolumn}
\usepackage{bm}
\usepackage{physics}
\usepackage{slashed}
\usepackage{xcolor}

\usepackage[all]{xy}

\usepackage{amscd}

\usepackage[
 bookmarks=true,
 bookmarksnumbered=true,
 colorlinks=false
]{hyperref}

\theoremstyle{definition}

\theoremstyle{plain}
\newtheorem{theorem}{Theorem}[section]

\newtheorem{corollary}[theorem]{Corollary}

\newtheorem{proposition}[theorem]{Proposition}

\theoremstyle{remark}

\begin{document}

\title{
Beyond Robustness of Magic: Quantifying Quantum State Convertibility
}

\author{Satoshi Kanno}
\email{satoshi.kanno06@g.softbank.co.jp}
\affiliation{
Quantum Information Technology Department,
Quantum Technology Division,
Product Research and Development Division,
SoftBank Corp.,
1-7-1 Kaigan, Minato-ku,
Tokyo 105-7529, Japan
}

\author{Yoshi-aki Shimada}
\email{yoshiaki.shimada01@g.softbank.co.jp}
\affiliation{
Quantum Information Technology Department,
Quantum Technology Division,
Product Research and Development Division,
SoftBank Corp.,
1-7-1 Kaigan, Minato-ku,
Tokyo 105-7529, Japan
}



\begin{abstract}
    In fault-tolerant quantum computing, magic states provide essential resources for implementing non-Clifford operations. The robustness of magic (RoM) quantifies the magic contained in a quantum state and is closely related to the cost of classical simulation. However, knowing how much magic two states contain is not sufficient to determine whether one can be converted into the other using free operations. In this work, we develop a geometric framework in which RoM is represented as half the distance to the set of free states, and extend this framework to state conversion. We introduce a directed conversion deficiency that measures how far a target state lies from the states freely reachable from a given input. This quantity vanishes exactly when free conversion is possible and, beyond being a mathematically well-defined measure, has an operational interpretation in terms of stabilizer quasiprobability. Furthermore, our framework provides lower bounds on the additional magic states required for conversion that can be strictly stronger than conventional RoM-based bounds. We demonstrate this distinction analytically by constructing a continuous family of mixed single-qubit states with identical RoM for which free conversion is possible in one direction but impossible in the reverse. Our results reveal that the usefulness of magic for quantum state conversion depends not only on its amount but also on the specific transformation being performed.
\end{abstract}

\maketitle



\section{Introduction}
\label{sec:introduction}

Quantum computational advantage cannot be achieved by stabilizer operations alone
(see, e.g., Refs.~\cite{gottesman1998heisenberg,aaronson2004improved,bravyi2005universal,anders2006fast,dehaene2003clifford,nest2008classical,jozsa2013classical,bravyi2016improved,bennink2017unbiased,bravyi2019simulation,howard2014contextuality,gidney2021stim}),
and non-stabilizerness, or \emph{magic}, is therefore a central resource for quantum computation.
In fault-tolerant quantum computation (FTQC), in particular, non-Clifford operations are
commonly implemented by consuming magic states together with stabilizer operations such as
Clifford unitaries and Pauli measurements.
Quantifying magic is thus relevant not only to the computational power of quantum states,
but also to the physical resources required to realize fault-tolerant computations.

Among the many magic monotones
\cite{veitch2014resource,veitch2012negative,wang2020efficiently,
seddon2021quantifying,wang2019quantifying,leone2022stabilizer,haug2023stabilizer,warmuz2025magic,sonya2025nonstabilizerness,bravyi2019simulation,heimendahl2021stabilizer,hahn2022quantifying,bu2025quantum,bu2025stabilizer,dowling2025magic,turkeshi2025magic},
the robustness of magic (RoM)
\cite{howard2017application,hamaguchi2024handbook,seddon2019quantifying,sarkar2020characterization,goto2022probing,saxena2022quantifying}
is distinguished by its direct operational interpretations.
In stabilizer-quasiprobability simulation, RoM determines the minimum quasiprobability
negativity, or equivalently the minimum total signed stabilizer weight, and thereby controls
the sampling overhead of the corresponding Monte Carlo methods
\cite{heinrich2019robustness,howard2017application}.
RoM also yields lower bounds on the number of nonstabilizer resource states required for
state preparation
\cite{howard2017application,Beverland_2020}.
Its usefulness therefore lies not only in assigning a scalar amount of magic to a state,
but also in connecting that scalar quantity to concrete operational tasks.

For state conversion, however, the total amount of magic is not by itself sufficient.
Let $\rho$ be an available input state and $\sigma$ a desired target state.
Monotonicity of RoM implies that
\begin{equation}
    R(\rho) < R(\sigma)
\end{equation}
rules out a free transformation $\rho \to \sigma$.
The converse is false in general: even when
\begin{equation}
    R(\rho) \geq R(\sigma),
\end{equation}
the target need not be freely reachable from the input.
Thus, the amount of magic contained in a state and the usefulness of that magic for
producing a specified target are distinct resource-theoretic questions.

State convertibility under stabilizer-preserving operations is itself well studied\cite{Ahmadi_2018,Zanoni_2025,saxena2022quantifying,heimendahl2022axiomatic,seddon2019quantifying,leone2024stabilizer,liu2019one,fang2020no,regula2020benchmarking,regula2021fundamental}.
Exact conversion can be characterized using semidefinite programming and complete families
of magic monotones~\cite{Ahmadi_2018}.
More generally, quantum resource theories admit conversion measures constructed from the
distance between a target state and the set of states freely reachable from a given input
state~\cite{Zanoni_2025}.
Consequently, our aim is neither to introduce a new yes/no criterion for magic-state
convertibility nor to claim the general notion of distance to a reachable set as new.

To address this question, we adopt a geometric viewpoint
on magic in quantum state space. Viewing quantum states
as points in state space, stabilizer states form a
particular convex subset. The robustness of magic (RoM)
can then be understood as a resource quantity that
measures how far a given state lies from this free-state
set. In this work, we introduce a distance calibrated
by stabilizer states based on the dual formulation of
RoM and show that RoM is exactly represented by the
distance to the free-state set. This geometric
representation is closely related to the base-norm
structure of existing convex resource theories and
makes explicit the geometry inherent in RoM.

However, when considering state conversion, the
distance to the free-state set alone is not sufficient.
What matters is where the states reachable from a
given input under free operations are located in state
space. Even if two states are at the same distance
from the free-state set, it does not necessarily
follow that one can be freely converted into the other.
We therefore replace the fixed free-state set with
the set of states freely reachable from a given input
and define the directed conversion deficiency
$C_{\mathcal F}(\rho\to\sigma)$ as the distance from
this reachable set to the target state. This quantity
provides a geometric measure of how closely the target
can be approached by free operations. It vanishes
if and only if an exact free conversion is possible,
whereas a strictly positive value indicates that
the target cannot be reached exactly. Thus, while
RoM quantifies the resource content of an individual
state, $C_{\mathcal F}$ captures the directed geometric
relation between an input state and a target state.

An important feature of this geometric construction
is that the underlying distance is not chosen
arbitrarily, but arises naturally from the dual
structure of RoM. Consequently, the directed
conversion deficiency has an operational
interpretation in terms of stabilizer
quasiprobability, rather than being merely an
abstract state-space distance. Specifically, after
optimal free processing of the input, the minimum
total absolute stabilizer weight required to
represent the remaining difference from the target
as a signed linear combination of stabilizer states
is exactly $2C_{\mathcal F}$. When the freely
processed reference state is also included, the
minimum conditional total quasiprobability weight
is given by
\begin{equation}
\Gamma_{\mathrm{conv}}(\rho\to\sigma)
=
1+2C_{\mathcal F}(\rho\to\sigma).
\end{equation}
Thus, the geometric distance directly quantifies
a conversion-specific quasiprobability burden
that cannot be eliminated by free processing.
This extends the relation between RoM and
stabilizer quasiprobability from individual states
to state conversion. However, this burden is
conditional on the conversion task and should
not be identified with the total classical
simulation complexity of the entire process.

We further extend this framework to resource-assisted
conversions involving additional magic states.
From a geometric perspective, supplying additional
resources enlarges the set of freely reachable
states and can change its distance from the target.
Using this viewpoint, we derive lower bounds on
the number of additional resource states required
for a given conversion. These bounds subsume
the conventional RoM-based bounds and can detect
conversion obstructions that cannot be identified
by comparing RoM values alone. The resulting
framework therefore enables resource states to
be evaluated not only in terms of their total
magic content, but also in terms of their
usefulness for producing a specified target state.

Finally, we demonstrate the significance of this
directed geometry in the simplest setting of a
single qubit. Specifically, we analytically
construct a continuous family of strictly mixed
states that have the same distance from the
free-state set, and hence identical RoM, but
admit free conversion in one direction while
the reverse conversion is impossible.
We explicitly realize the forward conversion
using a randomized Clifford channel and
evaluate the reverse conversion deficiency
exactly. This result demonstrates that even
when two states contain the same amount of
magic, their usefulness as resources can
differ depending on their positions in state
space and their reachability under free
operations.

\paragraph{Main contributions.}
The contributions of this work can be summarized as follows:
\begin{enumerate}
    \item
    We identify the RoM-induced state-space geometry that serves as the reference geometry
    for the subsequent conversion problem.

    \item
    We lift this geometry to a directed deficiency relative to the freely reachable set and
    show that its zero set exactly characterizes free convertibility.

    \item
    We establish the exact relation between the directed deficiency and the minimum signed
    stabilizer weight required to represent the conversion residual.

    \item
    We show that the resulting resource-assisted bounds contain, and can strictly
    strengthen, the conventional lower bounds obtained from RoM alone.

    \item
    We construct an analytical continuous family of equal-RoM mixed single-qubit states
    exhibiting asymmetric free convertibility and evaluate the nonzero conversion
    deficiency in closed form.
\end{enumerate}

Taken together, these results extend three structures associated with RoM---its
resource-adapted geometry, its stabilizer-quasiprobability interpretation, and its
magic-state resource bounds---from the quantification of resource content in an individual
state to the directed task of producing a specified target.
The resulting framework separates the question of \emph{how much} magic an input contains
from the question of \emph{how much of that resource is usable for a particular conversion}.

\section{A Brief Introduction to Noncommutative Geometry}

In this section, we briefly introduce the basic ideas of noncommutative metric geometry\cite{rieffel1999metrics,rieffel2004gromov,rieffel2003compact} that will be used throughout this work.
For the purposes of the present study, we do not require the full axiomatic framework of noncommutative metric geometry, such as the theory of compact quantum metric spaces.
Instead, we focus on the basic framework in which a suitable seminorm or norm on observables induces a dual distance on the state space.
This construction will serve as the starting point for introducing a distance adapted to the stabilizer resource theory in the following section.

We begin with an ordinary metric space $(X,d)$.
For a function $f:X\to\mathbb{C}$, its Lipschitz seminorm is defined by
\begin{equation}
\operatorname{Lip}(f)
:=
\sup_{\substack{x,y\in X\\x\neq y}}
\frac{|f(x)-f(y)|}{d(x,y)}.
\end{equation}
Using this seminorm, one can define a distance between two probability measures $\mu$ and $\nu$ on $X$ by
\begin{equation}
d_{\mathrm{Lip}}(\mu,\nu)
:=
\sup_{\operatorname{Lip}(f)\leq 1}
\left|
\int_X f\,d\mu
-
\int_X f\,d\nu
\right|.
\end{equation}
This is known as the Kantorovich--Rubinstein dual formulation.
In particular, for the Dirac measures $\delta_x$ and $\delta_y$ associated with two points $x,y\in X$, one has
\begin{equation}
d_{\mathrm{Lip}}(\delta_x,\delta_y)=d(x,y).
\end{equation}
Thus, the original distance between points can be reconstructed by maximizing the difference of expectation values over suitably normalized functions\cite{Landi:1997sh,Connes:1994yd,connes1985non,connes2000short}.

This idea admits a natural extension to the state space of a $C^*$-algebra.
In the commutative case, the states of the algebra $C(X)$ can be identified with probability measures on $X$.
The construction above may therefore be regarded as defining a distance between states from a Lipschitz seminorm on functions.
Noncommutative metric geometry extends this idea to noncommutative algebras by replacing functions with self-adjoint observables.

More specifically, let $\mathcal{A}$ be a unital $C^*$-algebra, and denote its self-adjoint part by
\begin{equation}
\mathcal{A}_{\mathrm{sa}}
:=
\{a\in\mathcal{A}\mid a=a^\dagger\}.
\end{equation}
Suppose that a seminorm
\begin{equation}
L:\mathcal{A}_{\mathrm{sa}}\to[0,\infty]
\end{equation}
is given on the self-adjoint elements.
Then a distance between two states $\varphi$ and $\psi$ can be defined by
\begin{equation}
d_L(\varphi,\psi)
:=
\sup_{\substack{a\in\mathcal{A}_{\mathrm{sa}}\\L(a)\leq 1}}
|\varphi(a)-\psi(a)|.
\end{equation}

For a finite-dimensional quantum system, states can be represented by density operators.
In this case, the distance takes the form
\begin{equation}
d_L(\rho,\sigma)
=
\sup_{\substack{A=A^\dagger\\L(A)\leq 1}}
\left|
\operatorname{Tr}\bigl[(\rho-\sigma)A\bigr]
\right|.
\end{equation}

The important point is that the geometry induced on the state space is determined by the choice of $L$, which specifies how the size of an observable is measured.
The condition
\begin{equation}
L(A)\leq 1
\end{equation}
defines the admissible set of test observables used to distinguish two states.
Consequently, different choices of which observables are regarded as having unit scale lead to different metric structures on the state space.
In the classical setting, choosing $L=\operatorname{Lip}$ yields a Kantorovich-type distance.
In a quantum setting, one may instead choose $L$ according to the physical structure or resource theory of interest.

In general, the quantity $d_L$ defined in this way may be an extended pseudometric.
That is, it need not distinguish all distinct states, and its finiteness is not guaranteed without appropriate assumptions.
Therefore, when such a construction is used as a genuine distance on the state space, one must verify properties such as nondegeneracy and finiteness.

\section{Robustness of Magic and Stabilizer Quasiprobability Simulation}

In this section, we review the robustness of magic (RoM) and its main operational meanings. In particular, we show that RoM quantifies the sign problem arising in classical Monte Carlo simulation based on stabilizer quasiprobability representations and controls the corresponding sampling overhead. We further show that RoM provides lower bounds on the number of magic resource states required for state preparation.

Let the $n$-qubit Hilbert space be
\[
\mathcal H_n=(\mathbb C^2)^{\otimes n},
\]
and let
\[
\mathcal A_n=\operatorname{End}(\mathcal H_n)
\]
denote the full matrix algebra. In finite dimensions, a state can be identified with a density operator $\rho\geq0$ satisfying $\operatorname{Tr}\rho=1$.

We define the set of pure stabilizer states by
\[
\operatorname{Stab}^{\rm pure}_n
=
\left\{
|\phi\rangle\langle\phi|
\;\middle|\;
|\phi\rangle=U|0^n\rangle,\;
U\in\operatorname{Cliff}_n
\right\},
\]
and define the stabilizer free-state set as its convex hull,
\[
\mathcal F:=\operatorname{Stab}_n
=
\operatorname{conv}
\bigl(
\operatorname{Stab}^{\rm pure}_n
\bigr).
\]
The class of free operations is defined by
\[
\operatorname{Free}
=
\left\{
\Lambda\ {\rm CPTP}
\;\middle|\;
\Lambda(\mathcal F)\subseteq\mathcal F
\right\}.
\]
Thus, a free operation does not generate magic from a stabilizer input.

In general, a map $M:S(\mathcal A_n)\longrightarrow\mathbb R_{\geq0}$
is called a magic monotone if it satisfies
\begin{enumerate}

\item[(i)] \textit{Non-negativity:}
\[
M(\rho)\geq 0.
\]

\item[(ii)] \textit{Faithfulness on the free set:}
\[
M(\rho)=0
\qquad (\rho\in\mathcal F).
\]

\item[(iii)] \textit{Monotonicity:}
\[
M(\Lambda(\rho))\leq M(\rho)
\]
for all $\Lambda\in\operatorname{Free}$.

\item[(iv)] \textit{Convexity:}
For $0\leq p\leq 1$,
\[
\begin{aligned}
M\!\left(p\rho_1+(1-p)\rho_2\right)
\quad\leq
pM(\rho_1)+(1-p)M(\rho_2).
\end{aligned}
\]

\end{enumerate}
RoM is a standard example of such a magic monotone.

In this work, we define the RoM of a state $\rho$ by
\[
R(\rho)
=
\inf
\left\{
t\geq0
\;\middle|\;
\rho=(1+t)\sigma-t\tau,\quad
\sigma,\tau\in\mathcal F
\right\}.
\]
In the magic-state resource-theory literature, it is also common to define the robustness of magic as the minimum $\ell_1$-weight $D(\rho)$ of a signed stabilizer decomposition, in which case $D(\rho)=1$ for free states. In this work, in accordance with our convention that a magic monotone vanishes on the free-state set, we refer to $R(\rho)=(D(\rho)-1)/2$ as the RoM. The two conventions are equivalent up to a simple affine transformation.
This definition measures the minimum negative weight required to express a nonstabilizer state as a signed combination of stabilizer states.

More generally, for a signed stabilizer decomposition
\[
\rho=\sum_i x_i\sigma_i,
\qquad
\sigma_i\in\mathcal F,
\]
we define
\[
D(\rho)
=
\inf
\left\{
\sum_i|x_i|
\;\middle|\;
\rho=\sum_i x_i\sigma_i
\right\}.
\]
Then
\[
D(\rho)=1+2R(\rho)
\]
holds. Indeed, since $\sum_i x_i=1$, if $p$ denotes the total positive weight and $n$ the total magnitude of the negative weight, then
\[
p-n=1,
\qquad
p+n=1+2n.
\]
Therefore, minimizing the negative weight is equivalent to minimizing the $\ell_1$-norm of the coefficients. Hence, $R(\rho)$ gives the minimum quasiprobability negativity, while $D(\rho)$ gives the minimum total quasiprobability weight.

We next consider stabilizer quasiprobability simulation. Here, a stabilizer process $\mathcal E$ is a process composed of stabilizer-state and stabilizer-ancilla preparation, Clifford unitaries, Pauli measurements, classical feed-forward, probabilistic mixing, and subsystem discarding. Such a process can be efficiently simulated classically when acting on a stabilizer input. Therefore,
\[
\text{stabilizer state}
+
\text{stabilizer process}
\]
does not by itself constitute the essential difficulty in classical simulation; the difficulty arises from the presence of a magic input. The stabilizer processes considered here should be distinguished from the more general stabilizer-preserving operations used later in the discussion of state conversion.

Suppose that a magic state $\rho$ is decomposed as
\[
\rho=\sum_i x_i\sigma_i,
\qquad
\gamma:=\sum_i|x_i|.
\]
Consider a stabilizer process $\mathcal E$ followed by the measurement of a Pauli observable $P$, and suppose that we wish to estimate
\[
\mu_\rho
=
\operatorname{Tr}[P\mathcal E(\rho)].
\]

In each Monte Carlo trial, choose $i$ according to
\[
p_i=\frac{|x_i|}{\gamma},
\]
sample $m\in\{-1,+1\}$ from the stabilizer simulation of $\sigma_i$, and define
\[
Y
=
\gamma\,\operatorname{sgn}(x_i)m.
\]
Then
\[
\mathbb E[Y]=\mu_\rho,
\qquad
|Y|\leq\gamma,
\]
so $Y$ is an unbiased estimator.

If all coefficients are nonnegative, then $\gamma=1$, and the state can be simulated as an ordinary probabilistic mixture. By contrast, for a magic state, negative coefficients are generally required and $\gamma>1$.

Using $N$ independent samples, define
\[
\widehat\mu_\rho
=
\frac1N\sum_{j=1}^N Y_j.
\]
Hoeffding's inequality gives
\[
\Pr
\left[
|\widehat\mu_\rho-\mu_\rho|
\geq\epsilon
\right]
\leq
2\exp
\left(
-\frac{N\epsilon^2}{2\gamma^2}
\right).
\]
Therefore, in order to guarantee additive error $\epsilon$ with failure probability at most $\delta$, it is sufficient to take
\[
N
\geq
\frac{2\gamma^2}{\epsilon^2}
\log\frac2\delta.
\]
For an optimal stabilizer decomposition,
\[
\gamma_{\min}=D(\rho)=1+2R(\rho),
\]
and hence
\[
N
\geq
\frac{
2(1+2R(\rho))^2
}{
\epsilon^2
}
\log\frac2\delta.
\]
Thus, $1+2R(\rho)$ gives the minimum stabilizer quasiprobability weight, while its square controls the corresponding sampling overhead.

This operational interpretation should, however, be understood specifically within Monte Carlo simulation based on stabilizer quasiprobability representations. It does not imply that RoM provides a universal lower bound on the computational complexity of every possible classical simulation algorithm.

RoM also provides lower bounds on the number of magic resource states required for state preparation. Let $\phi$ be a fixed magic resource state, and suppose that
\[
\sigma
=
\Lambda
\left(
\rho\otimes\phi^{\otimes k}
\right)
\]
can be realized by a free operation $\Lambda$. By monotonicity of $D$ and its submultiplicativity under tensor products,
\[
D(\rho\otimes\tau)
\leq
D(\rho)D(\tau),
\]
we obtain the necessary condition
\[
D(\sigma)
\leq
D(\rho)D(\phi)^k.
\]

If $D(\phi)>1$, this yields
\[
k
\geq
\max
\left\{
0,
\left\lceil
\frac{
\log D(\sigma)-\log D(\rho)
}{
\log D(\phi)
}
\right\rceil
\right\}.
\]
In particular, if the input state $\rho$ is free, then $D(\rho)=1$, and the bound reduces to
\[
k
\geq
\left\lceil
\frac{\log D(\sigma)}
{\log D(\phi)}
\right\rceil.
\]
This is the standard RoM-based lower bound on the number of magic resource states required to generate the target state.

Thus, the two main operational meanings of RoM can be summarized as the sampling overhead in stabilizer quasiprobability simulation and the resource lower bound in magic-state synthesis. The former quantifies the sign problem arising from quasiprobability negativity, while the latter provides a necessary condition for whether the available input contains a sufficient total amount of magic to generate the target state.

\section{Robustness of Magic as a Geometric Distance}

In the previous section, we saw that the robustness of magic (RoM) quantifies the negativity appearing in stabilizer quasiprobability representations and is directly related to both the sampling overhead of classical Monte Carlo simulation and lower bounds on magic-state resources. In this section, we introduce a state-space geometry adapted to the stabilizer resource theory through the dual structure of RoM.

Specifically, we define an observable norm $L_{\mathcal F}$ calibrated by the stabilizer free-state set $\mathcal F$, together with the corresponding state-space distance $d_{\mathcal F}$, and establish the central identity
\begin{equation}
R(\rho)
=
\frac{1}{2}
\operatorname{dist}_{\mathcal F}(\rho,\mathcal F).
\end{equation}
Thus, the present construction does not introduce a new magic monotone independent of RoM. Rather, it makes explicit the state-space geometry already encoded in the dual structure of RoM. This distance representation will later provide the starting point for extending state-level resource quantification to directed state conversion, where the fixed free-state set $\mathcal F$ is replaced by the set $\mathcal O(\rho)$ of states freely reachable from a given input state $\rho$.

\subsection{Dual Witness Formulation of RoM}

Recall that
\begin{equation}
D(\rho)=1+2R(\rho)
\end{equation}
is the minimum $\ell_1$-norm over signed stabilizer decompositions. Its dual optimization  problem\cite{howard2017application,hamaguchi2024handbook} is
\begin{equation}
1+2R(\rho)
=
\sup
\left\{
\operatorname{Tr}(\rho W)
\;\middle|\;
W=W^\dagger,\;
|\operatorname{Tr}(\omega W)|\leq 1
\quad
\forall\,\omega\in\mathcal F
\right\}.
\end{equation}
Hence, the feasible set in the dual formulation of RoM is characterized by
\begin{equation}
\sup_{\omega\in\mathcal F}
|\operatorname{Tr}(\omega W)|
\leq 1.
\end{equation}
We interpret this condition as a normalization of observables relative to the stabilizer free-state set and use it to define a resource-adapted observable norm.

\subsection{Stabilizer-Calibrated State-Space Distance}

For a Hermitian operator $A$, define
\begin{equation}
L_{\mathcal F}(A)
:=
\sup_{\omega\in\mathcal F}
|\operatorname{Tr}(\omega A)|.
\end{equation}

To bring the construction closer to the usual formulation of noncommutative metric geometry, one may quotient out the scalar direction and instead introduce the seminorm
\[
\widetilde{L}_{\mathcal F}(A)
:=
\inf_{c\in\mathbb{R}}
L_{\mathcal F}(A-cI),
\]
which vanishes on scalar multiples of the identity. This modification does not change the induced distance between normalized quantum states, since
\[
\operatorname{Tr}\!\left[(\rho-\sigma)cI\right]=0
\]
for any pair of states $\rho$ and $\sigma$. Hence, the state-space metric induced by $\widetilde{L}_{\mathcal F}$ coincides with $d_{\mathcal F}$ defined below. For notational simplicity, we therefore work directly with the norm $L_{\mathcal F}$ throughout this paper.

Since $\mathcal F$ is the convex hull of pure stabilizer states and the map
\[
\omega\longmapsto\operatorname{Tr}(\omega A)
\]
is affine, the optimization may be restricted to the extreme points of $\mathcal F$:
\begin{equation}
L_{\mathcal F}(A)
=
\max_{\phi\in\operatorname{Stab}^{\mathrm{pure}}_n}
|\operatorname{Tr}(\phi A)|.
\end{equation}
Thus, $L_{\mathcal F}$ measures the size of an observable in units calibrated by stabilizer states. In contrast to the ordinary operator norm, which is calibrated by the full quantum state space, $L_{\mathcal F}$ uses only the free-state set and thereby induces a geometry adapted specifically to the resource theory of magic.

Using $L_{\mathcal F}$, we define the distance between two quantum states $\rho$ and $\sigma$ by
\begin{equation}
d_{\mathcal F}(\rho,\sigma)
:=
\sup_{\substack{A=A^\dagger\\
L_{\mathcal F}(A)\leq 1}}
\left|
\operatorname{Tr}
\left[
(\rho-\sigma)A
\right]
\right|.
\end{equation}
The distance $d_{\mathcal F}$ measures the largest expectation-value difference between two states that can be detected by observables whose magnitude is normalized relative to the stabilizer free-state set.

For comparison, let $\mathcal S$ denote the full quantum state space and define
\begin{equation}
L_{\mathcal S}(A)
:=
\sup_{\omega\in\mathcal S}
|\operatorname{Tr}(\omega A)|.
\end{equation}
For Hermitian $A$,
\begin{equation}
L_{\mathcal S}(A)=\|A\|_\infty,
\end{equation}
and the corresponding dual distance is
\begin{equation}
d_{\mathcal S}(\rho,\sigma)
=
\|\rho-\sigma\|_1.
\end{equation}
Since $\mathcal F\subset\mathcal S$,
we have
$L_{\mathcal F}(A)
\leq
\|A\|_\infty$.
Thus, $d_{\mathcal F}$ compares quantum states using a generally larger class of test observables than the trace norm and defines a stabilizer-adapted state-space geometry.


The quantity $d_{\mathcal F}$ is in fact a genuine metric in finite dimensions.

\begin{proposition}
For a finite-dimensional $n$-qubit system, $L_{\mathcal F}$ is a norm on the real vector space of Hermitian operators, and $d_{\mathcal F}$ is a finite metric on the quantum state space. In particular,
\begin{equation}
d_{\mathcal F}(\rho,\sigma)=0
\quad\Longleftrightarrow\quad
\rho=\sigma.
\end{equation}
Moreover, there exists a finite constant $\alpha_n>0$ such that
\begin{equation}
d_{\mathcal F}(\rho,\sigma)
\leq
\alpha_n
\|\rho-\sigma\|_1.
\end{equation}
\end{proposition}
The proof is given in Appendix~\ref{app:metric-properties}. The essential observation is that stabilizer states span the full real vector space of Hermitian operators, which guarantees the nondegeneracy of $L_{\mathcal F}$.


The stabilizer-calibrated distance is compatible with free processing.

\begin{proposition}
Let $\Lambda$ be a free operation satisfying
\begin{equation}
\Lambda(\mathcal F)\subseteq\mathcal F.
\end{equation}
Then
\begin{equation}
d_{\mathcal F}
\bigl(
\Lambda(\rho),\Lambda(\sigma)
\bigr)
\leq
d_{\mathcal F}(\rho,\sigma).
\end{equation}
\end{proposition}
Thus, applying the same free operation to two states cannot increase their stabilizer-calibrated distance. The proof is given in Appendix~\ref{app:contractivity}.

\subsection{Exact Distance Representation of RoM}

We define the distance from a state $\rho$ to the stabilizer free-state set $\mathcal F$ by
\begin{equation}
\operatorname{dist}_{\mathcal F}
(\rho,\mathcal F)
:=
\inf_{\omega\in\mathcal F}
d_{\mathcal F}(\rho,\omega).
\end{equation}
The central result of this section is the following.

\begin{theorem}
For every quantum state $\rho$,
\begin{equation}
\operatorname{dist}_{\mathcal F}
(\rho,\mathcal F)
=
2R(\rho).
\end{equation}
\end{theorem}

The proof is given in Appendix~\ref{app:rom-distance}.
This result shows that RoM is not merely a convex-optimization quantity defined through signed stabilizer decompositions or dual witness optimization. In the geometry induced by $L_{\mathcal F}$, RoM is exactly the distance from a quantum state to the stabilizer free-state set.
It is important to emphasize that this metric construction does not introduce a new single-state magic monotone independent of RoM. Rather, the equality above makes explicit the resource-adapted geometry already encoded in the dual witness structure of RoM.


The distance representation immediately implies a stability relation for RoM.
\begin{corollary}
For arbitrary quantum states $\rho$ and $\sigma$,
\begin{equation}
|R(\rho)-R(\sigma)|
\leq
\frac{1}{2}
d_{\mathcal F}(\rho,\sigma).
\end{equation}
\end{corollary}
Thus, if two states are close in the stabilizer-calibrated geometry, their RoM values must also be close. The metric $d_{\mathcal F}$ therefore provides a natural resource-adapted scale that directly controls variations in magic. The proof follows from the general $1$-Lipschitz property of the distance-to-a-set function and is given in Appendix~\ref{app:rom-stability}.

\section{Directed Geometry of Magic-State Conversion}

For a general state-conversion problem $\rho\longrightarrow\sigma$,
however, the input state $\rho$ may already contain magic. The relevant question is therefore not merely how far $\sigma$ lies from $\mathcal F$, but how closely one can approach $\sigma$ after optimally processing $\rho$ using free operations.

In this section, we therefore replace the fixed reference set $\mathcal F$
appearing in the geometric representation of RoM by the input-dependent set $\mathcal O(\rho)$
of states freely reachable from $\rho$. This extends the state-level geometry underlying RoM to a geometry of directed state conversion.

The construction below can equivalently be understood in the convex-geometric language of atomic gauges and base norms developed in Ref. \cite{regula2018convex}. In particular, the symmetrized free-state set $F\cup (-F)$ generates a base norm on the Hermitian operator space, while the observable norm $L_F$ introduced below is its dual norm. We use this structure to formulate the geometry directly on the quantum state space and subsequently extend it to directed state conversion.

\subsection{Freely Reachable Sets and Directed Conversion Deficiency}

With respect to the stabilizer free-state set $\mathcal F$, we define the class of stabilizer-preserving operations (SPO) by
\begin{equation}
\mathrm{SPO}
:=
\left\{
\Lambda
\;\middle|\;
\Lambda\text{ is CPTP and }
\Lambda(\mathcal F)\subseteq\mathcal F
\right\}.
\end{equation}
The SPO considered here should be distinguished from the physically implementable stabilizer processes used in the previous discussion of classical simulation. The latter are explicitly constructed from Clifford operations, Pauli measurements, stabilizer ancillas, and related operations, whereas an SPO is defined solely by the resource-theoretic condition $\Lambda(\mathcal F)\subseteq\mathcal F$.
For an input state $\rho$, we define its freely reachable set by
\begin{equation}
\mathcal O(\rho)
:=
\left\{
\Lambda(\rho)
\;\middle|\;
\Lambda\in\mathrm{SPO}
\right\}.
\end{equation}

Motivated by the geometric representation $R(\sigma)
=
\frac{1}{2}
\operatorname{dist}_{\mathcal F}(\sigma,\mathcal F)$,
we replace $\mathcal F$ by $\mathcal O(\rho)$ and define the \emph{directed conversion deficiency} by
\begin{equation}
C_{\mathcal F}(\rho\to\sigma)
:=
\frac{1}{2}
\operatorname{dist}_{\mathcal F}
\bigl(
\sigma,\mathcal O(\rho)
\bigr).
\end{equation}
Equivalently,
\begin{equation}
C_{\mathcal F}(\rho\to\sigma)
=
\frac{1}{2}
\inf_{\Lambda\in\mathrm{SPO}}
d_{\mathcal F}
\bigl(
\Lambda(\rho),\sigma
\bigr).
\end{equation}

Thus, $C_{\mathcal F}(\rho\to\sigma)$ measures the minimum stabilizer-calibrated residual error that remains relative to the target state $\sigma$ even after optimally processing the input state $\rho$ by free operations.

Its most fundamental property is the following.
\begin{proposition}
In finite dimensions,
\begin{equation}
C_{\mathcal F}(\rho\to\sigma)=0
\iff
\exists\,\Lambda\in\mathrm{SPO}
\text{ such that }
\Lambda(\rho)=\sigma.
\end{equation}
\end{proposition}
The proof is given in Appendix~\ref{app:zero-conversion}.
Thus, the zero set of $C_{\mathcal F}$ completely characterizes exact free convertibility, while $C_{\mathcal F}>0$ quantifies the conversion obstruction that cannot be removed by free processing as a continuous quantity.


The directed conversion deficiency satisfies the following basic properties.
\begin{proposition}
For arbitrary states $\rho$, $\sigma$, and $\tau$, and for any $\Gamma\in\mathrm{SPO}$,
\begin{equation}
C_{\mathcal F}(\rho\to\sigma)\geq0,
\qquad
C_{\mathcal F}(\rho\to\rho)=0.
\end{equation}
In general,
\begin{equation}
C_{\mathcal F}(\rho\to\sigma)
\neq
C_{\mathcal F}(\sigma\to\rho).
\end{equation}
Moreover, the directed triangle inequality
\begin{equation}
C_{\mathcal F}(\rho\to\tau)
\leq
C_{\mathcal F}(\rho\to\sigma)
+
C_{\mathcal F}(\sigma\to\tau)
\end{equation}
holds.
Under free preprocessing of the input,
\begin{equation}
C_{\mathcal F}
\bigl(
\Gamma(\rho)\to\sigma
\bigr)
\geq
C_{\mathcal F}(\rho\to\sigma),
\end{equation}
whereas under free postprocessing of the target,
\begin{equation}
C_{\mathcal F}
\bigl(
\rho\to\Gamma(\sigma)
\bigr)
\leq
C_{\mathcal F}(\rho\to\sigma).
\end{equation}
\end{proposition}

The proof is given in Appendix~\ref{app:conversion-properties}.
These properties show that $C_{\mathcal F}$ is not an ordinary symmetric state-space metric, but rather a directed quantity compatible with the free-conversion preorder. In particular,
$
\sigma\in\mathcal O(\rho)
$
does not generally imply
$
\rho\in\mathcal O(\sigma).
$
Consequently, one may have
\begin{equation}
C_{\mathcal F}(\rho\to\sigma)=0,
\qquad
C_{\mathcal F}(\sigma\to\rho)>0.
\end{equation}

\subsection{Relation to the Robustness of Magic}

These quantities satisfy the following relation.
\begin{theorem}
For arbitrary states $\rho$ and $\sigma$,
\begin{equation}
[R(\sigma)-R(\rho)]_+
\leq
C_{\mathcal F}(\rho\to\sigma)
\leq
R(\sigma),
\end{equation}
where
\begin{equation}
[x]_+:=\max\{x,0\}.
\end{equation}
\end{theorem}

The proof is given in Appendix~\ref{app:conversion-rom-bounds}.
The lower bound implies that if
$
R(\sigma)>R(\rho),
$
then necessarily
\begin{equation}
C_{\mathcal F}(\rho\to\sigma)>0,
\end{equation}
and exact free conversion is impossible. Thus, $C_{\mathcal F}$ contains the conventional conversion obstruction obtained from the monotonicity of RoM.
The converse, however, does not hold. Even if
$
R(\rho)\geq R(\sigma),
$
it may still be the case that
\begin{equation}
C_{\mathcal F}(\rho\to\sigma)>0.
\end{equation}
In particular, even when
$
R(\rho)=R(\sigma)
$
,one may have
\begin{equation}
C_{\mathcal F}(\rho\to\sigma)>0.
\end{equation}

RoM compares only the total amount of magic contained in the two states, whereas $C_{\mathcal F}$ evaluates the geometric relation between the target state $\sigma$ and the entire reachable set $\mathcal O(\rho)$
generated from the input state $\rho$. Hence, $C_{\mathcal F}$ can detect conversion obstructions that are invisible to scalar comparisons of RoM.
Conceptually, the distinction can be summarized as
\begin{equation}
R(\rho):
\quad
\text{How much magic does }\rho\text{ contain?}
\end{equation}
versus
\begin{equation}
C_{\mathcal F}(\rho\to\sigma):
\quad
\text{Can the magic in }\rho
\text{ be directed toward }\sigma?
\end{equation}

\section{Magic-State Conversion as a Classical-Simulation Bottleneck}

At the state level, $D(\rho)=1+2R(\rho)$
gives the minimum stabilizer quasiprobability weight of $\rho$, and its square controls the sampling overhead in stabilizer-quasiprobability-based Monte Carlo simulation.
In this section, we extend this quasiprobability interpretation to directed state conversion.

\subsection{Signed Stabilizer Norm of the Conversion Residual}

For a Hermitian operator$X$, we define
\begin{equation}
\|X\|_{\mathcal F}
:=
\sup_{\substack{A=A^\dagger\\
L_{\mathcal F}(A)\leq1}}
|\operatorname{Tr}(XA)|.
\end{equation}
For differences of quantum states,
\begin{equation}
\|\rho-\sigma\|_{\mathcal F}
=
d_{\mathcal F}(\rho,\sigma).
\end{equation}
This dual definition admits a primal representation in terms of signed stabilizer decompositions.

\begin{proposition}
\label{prop:signed-stabilizer-norm}
For every Hermitian operator $X$,
\begin{equation}
\|X\|_{\mathcal F}
=
\inf
\left\{
\sum_j|q_j|
\;\middle|\;
X=\sum_jq_j\omega_j,\quad
q_j\in\mathbb R,\quad
\omega_j\in\mathcal F
\right\}.
\end{equation}
In particular, if
$
\operatorname{Tr}(X)=0,
$
then
\begin{equation}
\|X\|_{\mathcal F}
=
2
\inf
\left\{
t\geq0
\;\middle|\;
X=t(\omega_+-\omega_-),\quad
\omega_\pm\in\mathcal F
\right\}.
\end{equation}
\end{proposition}

The proof is given in Appendix~\ref{app:residual-norm}.
For the state-conversion problem, let
\begin{equation}
X
=
\sigma-\Lambda(\rho).
\end{equation}
Since $\sigma$ and $\Lambda(\rho)$ are both normalized states,
$
\operatorname{Tr}(X)=0,
$
Therefore,
$
d_{\mathcal F}
\bigl(
\Lambda(\rho),\sigma
\bigr)
$
is exactly the minimum total absolute weight required to represent the residual
$
\sigma-\Lambda(\rho)
$
as a signed linear combination of stabilizer states.
Hence, $d_{\mathcal F}$ is not merely an abstract state-space distance. It admits the quasiprobability interpretation : minimum signed stabilizer weight required to correct the conversion residual.

\subsection{Conversion Quasiprobability Weight}

Let $\Lambda\in\mathrm{SPO}$
be a free operation and represent the target state as
\begin{equation}
\sigma
=
\Lambda(\rho)
+
\sum_jq_j\omega_j,
\qquad
\omega_j\in\mathcal F.
\end{equation}
Here, $\Lambda(\rho)$ is the freely processed reference state, while $\sum_jq_j\omega_j$
is the signed stabilizer correction required to reach the target.
Taking the trace gives $\sum_jq_j=0.$

For such a representation, we consider the conditional total quasiprobability weight
\begin{equation}
1+\sum_j|q_j|,
\end{equation}
where the leading coefficient $1$ corresponds to the reference branch. We define its minimum value by
\begin{align}
&\Gamma_{\mathrm{conv}}(\rho\to\sigma)\\ \nonumber
&:=
\inf
\left\{
1+\sum_j|q_j|
\;\middle|\;
\sigma
=
\Lambda(\rho)
+
\sum_jq_j\omega_j,\;
\Lambda\in\mathrm{SPO},\;
\omega_j\in\mathcal F
\right\}.
\end{align}

The central result of this section is the following.
\begin{theorem}
For every state-conversion problem $\rho\to\sigma$,
\begin{equation}
\Gamma_{\mathrm{conv}}(\rho\to\sigma)
=
1+2C_{\mathcal F}(\rho\to\sigma).
\end{equation}
\end{theorem}

The proof is given in Appendix~\ref{app:conversion-weight}.
Thus, $2C_{\mathcal F}(\rho\to\sigma)$
is the minimum signed stabilizer correction weight that remains necessary after optimal free processing, whereas $1+2C_{\mathcal F}(\rho\to\sigma)$
is the minimum conditional total quasiprobability weight including the reference branch.
Thus,
\begin{equation}
\Gamma_{\mathrm{conv}}(\rho\to\sigma)
=
1+2C_{\mathcal F}(\rho\to\sigma)
\end{equation}
give the conversion-level quasiprobability weights, respectively.

\subsection{Monte Carlo Interpretation}

We now connect the preceding quasiprobability representation to Monte Carlo sampling overhead.

For a Pauli observable $P$, consider the expectation-value difference between the target state and a freely processed input,
\begin{equation}
\Delta_P(\Lambda)
:=
\operatorname{Tr}(P\sigma)
-
\operatorname{Tr}
\left[
P\Lambda(\rho)
\right].
\end{equation}
Suppose that the conversion residual is decomposed as
\begin{equation}
\sigma-\Lambda(\rho)
=
\sum_jq_j\omega_j,
\qquad
\omega_j\in\mathcal F,
\end{equation}
and define
\begin{equation}
\gamma_{\mathrm{res}}
:=
\sum_j|q_j|.
\end{equation}

\begin{proposition}
The signed residual decomposition above yields an unbiased Monte Carlo estimator of $\Delta_P(\Lambda)$.
To estimate $\Delta_P(\Lambda)$ within additive error $\epsilon$ with failure probability at most $\delta$, it is sufficient to use
\begin{equation}
N_{\mathrm{res}}
\geq
\frac{
2\gamma_{\mathrm{res}}^2
}{
\epsilon^2
}
\log
\left(
\frac{2}{\delta}
\right)
\end{equation}
samples.
Optimizing over both the free operation and the residual decomposition gives
\begin{equation}
\gamma_{\mathrm{res}}^{\min}
=
2C_{\mathcal F}(\rho\to\sigma),
\end{equation}
and hence it is sufficient to take
\begin{equation}
N_{\mathrm{res}}
\geq
\frac{
8C_{\mathcal F}(\rho\to\sigma)^2
}{
\epsilon^2
}
\log
\left(
\frac{2}{\delta}
\right).
\end{equation}
\end{proposition}

The proof is given in Appendix~\ref{app:residual-monte-carlo}.
Therefore, if $C_{\mathcal F}(\rho\to\sigma)=0,$
there exists a free operation satisfying $\Lambda(\rho)=\sigma,$
so no residual correction is required and there is no sign problem associated with the conversion residual.
By contrast, if $C_{\mathcal F}(\rho\to\sigma)>0,$
no free processing can completely eliminate the residual, and a minimum total signed correction of
\begin{equation}
2C_{\mathcal F}(\rho\to\sigma)
\end{equation}
remains unavoidable.
In this sense, $C_{\mathcal F}(\rho\to\sigma)$
quantifies the \emph{conversion-specific sign problem} that cannot be removed by free processing.

\subsubsection{Conditional Reconstruction of the Target}

We next consider reconstruction of the target expectation value $\operatorname{Tr}(P\sigma)$
itself, rather than only the residual expectation value.
We assume that, for a chosen free operation $\Lambda$, sample access to Pauli measurement outcomes of $\Lambda(\rho)$
is already available. Such sample access may arise from a separate classical simulation, experimental measurement data, or oracle access.

\begin{proposition}
For a representation
\begin{equation}
\sigma
=
\Lambda(\rho)
+
\sum_jq_j\omega_j,
\end{equation}
one can construct a quasiprobability Monte Carlo estimator that combines the reference branch with the residual branches.
Let
\begin{equation}
\Gamma
=
1+\sum_j|q_j|
\end{equation}
be its total absolute weight. To achieve additive error $\epsilon$ with failure probability at most $\delta$, it is sufficient to use
\begin{equation}
N_{\mathrm{cond}}
\geq
\frac{
2\Gamma^2
}{
\epsilon^2
}
\log
\left(
\frac{2}{\delta}
\right)
\end{equation}
samples.
After optimization,
\begin{equation}
\Gamma
=
\Gamma_{\mathrm{conv}}(\rho\to\sigma)
=
1+2C_{\mathcal F}(\rho\to\sigma),
\end{equation}
and therefore
\begin{equation}
N_{\mathrm{cond}}
\geq
\frac{
2
\left(
1+2C_{\mathcal F}(\rho\to\sigma)
\right)^2
}{
\epsilon^2
}
\log
\left(
\frac{2}{\delta}
\right).
\end{equation}
\end{proposition}

The proof is given in Appendix~\ref{app:conditional-reconstruction}.
Thus, the quantity $1+2C_{\mathcal F}$ represents
the minimum conditional quasiprobability weight
and thereby quantifies the quasiprobability
bottleneck associated with state conversion.

\subsection{Scope of the Conversion-Simulation Interpretation}

Let
\begin{equation}
\rho
=
\sum_i x_i\phi_i,
\qquad
\sum_i |x_i|
=
D(\rho)
\end{equation}
be an optimal stabilizer quasiprobability decomposition of the input state.
For a free operation $\Lambda\in\mathrm{SPO}$, let
\begin{equation}
\sigma
=
\Lambda(\rho)
+
\sum_j q_j\omega_j
\end{equation}
be a conversion correction.
By linearity,
\begin{equation}
\sigma
=
\sum_i x_i\Lambda(\phi_i)
+
\sum_j q_j\omega_j.
\end{equation}

\begin{proposition}
Suppose that $\Lambda$ admits efficient classical sampling on each stabilizer input
$\phi_i$.
Then the above decomposition defines a conversion-respecting quasiprobability
simulation strategy.

For a fixed $\Lambda$, optimizing over the residual stabilizer decomposition gives
\begin{equation}
\inf_{\{q_j,\omega_j\}}
\sum_j |q_j|
=
d_{\mathcal F}
\bigl(
\Lambda(\rho),\sigma
\bigr),
\end{equation}
and hence the total absolute weight of the resulting representation is bounded by
\begin{equation}
D(\rho)
+
d_{\mathcal F}
\bigl(
\Lambda(\rho),\sigma
\bigr).
\end{equation}
Therefore, it is sufficient to use
\begin{equation}
N_{\mathrm{path}}
\geq
\frac{
2
\left[
D(\rho)
+
d_{\mathcal F}
\bigl(
\Lambda(\rho),\sigma
\bigr)
\right]^2
}{
\epsilon^2
}
\log
\left(
\frac{2}{\delta}
\right)
\end{equation}
samples.

In particular, if there exists a classically simulable
$\Lambda_\star\in\mathrm{SPO}$ that attains the optimization defining
$C_{\mathcal F}(\rho\to\sigma)$, so that
\begin{equation}
d_{\mathcal F}
\bigl(
\Lambda_\star(\rho),\sigma
\bigr)
=
2C_{\mathcal F}(\rho\to\sigma),
\end{equation}
then the total absolute weight is bounded by
\begin{equation}
D(\rho)
+
2C_{\mathcal F}(\rho\to\sigma),
\end{equation}
and consequently
\begin{equation}
N_{\mathrm{path}}
\geq
\frac{
2
\left[
D(\rho)
+
2C_{\mathcal F}(\rho\to\sigma)
\right]^2
}{
\epsilon^2
}
\log
\left(
\frac{2}{\delta}
\right)
\end{equation}
samples are sufficient.
Using
\begin{equation}
D(\rho)=1+2R(\rho),
\end{equation}
this can equivalently be written as
\begin{equation}
N_{\mathrm{path}}
\geq
\frac{
2
\left[
1+2R(\rho)
+
2C_{\mathcal F}(\rho\to\sigma)
\right]^2
}{
\epsilon^2
}
\log
\left(
\frac{2}{\delta}
\right).
\end{equation}
\end{proposition}

The proof is given in Appendix~\ref{app:full-path-simulation}.
This is a constructive upper bound for the particular conversion-respecting
simulation strategy described above and is not claimed to characterize the
optimal complexity of arbitrary classical simulation methods.
The quantity $C_{\mathcal F}$ determines the minimum residual signed stabilizer
weight over the full class of stabilizer-preserving operations, but the
operation attaining this minimum need not itself admit an efficient classical
simulation.
Moreover, additional cancellations between the state-level quasiprobability
decomposition and the residual correction may reduce the actual total variation
below the bound above.

\subsection{SPO and Algorithmic Simulatability}

The preceding discussion highlights an important distinction between the
resource-theoretic and algorithmic interpretations of the conversion
deficiency.
The definition
\begin{equation}
C_{\mathcal F}(\rho\to\sigma)
=
\frac{1}{2}
\inf_{\Lambda\in\mathrm{SPO}}
d_{\mathcal F}
\bigl(
\Lambda(\rho),\sigma
\bigr)
\end{equation}
involves optimization over the full class of stabilizer-preserving operations.
However, the condition
$\Lambda(\mathcal F)\subseteq\mathcal F$
is purely resource-theoretic and does not, by itself, imply that $\Lambda$
admits an efficient classical simulation on stabilizer inputs.
Thus, $C_{\mathcal F}$ should be interpreted as the
\emph{information-theoretic minimum signed stabilizer correction} achievable
by free processing, rather than as a general statement about the computational
cost of implementing the optimizing operation.

For a literal algorithmic classical-simulation interpretation, we therefore
introduce a subclass
\begin{equation}
\mathrm{SPO}_{\mathrm{sim}}
\subseteq
\mathrm{SPO}
\end{equation}
consisting of stabilizer-preserving operations that admit efficient classical
sampling on stabilizer inputs.
Restricting the optimization to this subclass, we define
\begin{equation}
C_{\mathcal F}^{\mathrm{sim}}
(\rho\to\sigma)
:=
\frac{1}{2}
\inf_{\Lambda\in\mathrm{SPO}_{\mathrm{sim}}}
d_{\mathcal F}
\bigl(
\Lambda(\rho),\sigma
\bigr).
\end{equation}
By the same signed-stabilizer decomposition argument as for the full SPO class,
the corresponding conversion quasiprobability weight is
\begin{equation}
\Gamma_{\mathrm{conv}}^{\mathrm{sim}}
(\rho\to\sigma)
=
1+
2C_{\mathcal F}^{\mathrm{sim}}
(\rho\to\sigma).
\end{equation}

Since
$\mathrm{SPO}_{\mathrm{sim}}\subseteq\mathrm{SPO}$,
the optimization is performed over a smaller set, and therefore
\begin{equation}
C_{\mathcal F}(\rho\to\sigma)
\leq
C_{\mathcal F}^{\mathrm{sim}}(\rho\to\sigma).
\end{equation}
Accordingly,
$C_{\mathcal F}(\rho\to\sigma)$
quantifies the intrinsic conversion bottleneck associated with the minimum
signed stabilizer correction allowed by the full resource theory, whereas
$C_{\mathcal F}^{\mathrm{sim}}(\rho\to\sigma)$
is the corresponding bottleneck when free processing is restricted to
operations admitting efficient classical sampling on stabilizer inputs.

If an optimizer for $C_{\mathcal F}$ belongs to
$\mathrm{SPO}_{\mathrm{sim}}$, then
\begin{equation}
C_{\mathcal F}^{\mathrm{sim}}(\rho\to\sigma)
=
C_{\mathcal F}(\rho\to\sigma),
\end{equation}
and the information-theoretic and algorithmic conversion bottlenecks coincide.
In general, however, the inequality may be strict.

\section{Stronger Magic-State Lower Bounds from Directed Conversion Geometry}

In this section, we first define the conversion deficiency in the presence of additional magic resource states. We then show that the standard RoM-based lower bound is automatically recovered from this quantity, while the conversion deficiency may remain strictly positive even after the RoM test becomes inconclusive. Thus, directed conversion geometry contains the standard RoM lower bound and can strengthen it strictly.

\subsection{Resource-Assisted Conversion}

We consider the general setting in which the input, resource, and target states may belong to different systems.

Let $A$ denote the system containing the initial input state $\rho$, let $R$ denote the system carrying an additional magic resource state $\phi$, and let $B$ denote the system containing the target state $\sigma$. We denote the corresponding stabilizer free-state sets by
\begin{equation}
\mathcal F_A,\qquad
\mathcal F_R,\qquad
\mathcal F_B.
\end{equation}
We define the class of stabilizer-preserving operations from system $A$ to system $B$ by
\begin{align}
&\mathrm{SPO}(A\to B)\\ \nonumber
&:=
\left\{
\Lambda:A\to B
\;\middle|\;
\Lambda\text{ is CPTP and }
\Lambda(\mathcal F_A)\subseteq\mathcal F_B
\right\}.
\end{align}
For an input state $\eta$ and a target state $\sigma$, define
\begin{equation}
C_{A\to B}(\eta\to\sigma)
:=
\frac{1}{2}
\inf_{\Lambda\in\mathrm{SPO}(A\to B)}
d_{\mathcal F_B}
\bigl(
\Lambda(\eta),\sigma
\bigr).
\end{equation}
Here, the distance is calibrated by $\mathcal F_B$ because the conversion residual belongs to the output system $B$.

Let $\phi$
be a fixed nonfree resource state, and suppose that $k\in\mathbb N_0$
copies of $\phi$ are available. We define $\eta_k
:=
\rho\otimes\phi^{\otimes k}.$
The resource-assisted conversion deficiency with $k$ copies of $\phi$ is then defined by
\begin{equation}
C_{\phi}^{(k)}(\rho\to\sigma)
:=
C_{A\otimes R^{\otimes k}\to B}
\left(
\rho\otimes\phi^{\otimes k}
\to\sigma
\right).
\end{equation}
Equivalently,
\begin{equation}
C_{\phi}^{(k)}(\rho\to\sigma)
=
\frac{1}{2}
\inf_{
\Lambda\in
\mathrm{SPO}
(A\otimes R^{\otimes k}\to B)
}
d_{\mathcal F_B}
\left(
\Lambda
\left(
\rho\otimes\phi^{\otimes k}
\right),
\sigma
\right).
\end{equation}
This quantity measures the minimum residual deficiency relative to the target after supplying $k$ copies of the magic resource state and optimizing over all free processing.
\begin{proposition}
In finite dimensions,
\begin{equation}
C_{\phi}^{(k)}(\rho\to\sigma)=0
\end{equation}
if and only if there exists
\begin{equation}
\Lambda
\in
\mathrm{SPO}
(A\otimes R^{\otimes k}\to B)
\end{equation}
such that $\Lambda
\left(
\rho\otimes\phi^{\otimes k}
\right)
=
\sigma.$
\end{proposition}

Therefore, $C_{\phi}^{(k)}(\rho\to\sigma)>0$
certifies that exact conversion is impossible using only $k$ copies of the resource state.
Using this zero-set characterization, we define the exact resource-state cost by
\begin{equation}
k_{\phi}^{\min}(\rho\to\sigma)
:=
\min
\left\{
k\in\mathbb N_0
\;\middle|\;
C_{\phi}^{(k)}(\rho\to\sigma)=0
\right\}.
\end{equation}
If exact conversion is impossible with any finite number of resource states, we adopt the convention $k_{\phi}^{\min}(\rho\to\sigma)=+\infty.$
Thus, $k_{\phi}^{\min}$ is not merely a lower bound but the exact resource-state cost for the specified class of free operations.

Furthermore, assuming that discarding a subsystem is a free operation, supplying additional resource states cannot increase the conversion deficiency.

\begin{proposition}
\begin{equation}
C_{\phi}^{(k+1)}(\rho\to\sigma)
\leq
C_{\phi}^{(k)}(\rho\to\sigma).
\end{equation}
Consequently, if for some $K$,
\begin{equation}
C_{\phi}^{(K)}(\rho\to\sigma)>0,
\end{equation}
then
\begin{equation}
k_{\phi}^{\min}(\rho\to\sigma)
\geq
K+1.
\end{equation}
\end{proposition}
The proof is given in Appendix~\ref{app:assisted-conversion}.

\subsection{Recovery and Strengthening of the RoM Lower Bound}

For each system $X$, define $D_X(\omega)
:=
1+2R_X(\omega)$
where $R_X$ denotes the robustness of magic relative to $\mathcal F_X$.
The quantity $D_X$ is monotone under stabilizer-preserving operations, so that $D_B(\Lambda(\eta))
\leq
D_A(\eta).$
It is also submultiplicative under tensor products:
\begin{equation}
D_{A\otimes R}(\rho\otimes\phi)
\leq
D_A(\rho)D_R(\phi).
\end{equation}
Therefore, if exact conversion using $k$ copies of $\phi$ is possible, the necessary condition
\begin{equation}
D_B(\sigma)
\leq
D_A(\rho)D_R(\phi)^k
\end{equation}
must hold.
When $D_R(\phi)>1,$
the standard RoM-based lower bound can be written as
\begin{equation}
k_{\phi}^{\mathrm{RoM}}(\rho\to\sigma)
:=
\max
\left\{
0,
\left\lceil
\frac{
\log D_B(\sigma)-\log D_A(\rho)
}{
\log D_R(\phi)
}
\right\rceil
\right\}.
\end{equation}
In particular, when $\rho$ is a free state,
\begin{equation}
D_A(\rho)=1,
\end{equation}
and this reduces to the usual RoM-based lower bound on the number of magic resource states.

The directed conversion deficiency automatically contains this RoM criterion.

\begin{theorem}
For every $k\in\mathbb N_0,$
we have
\begin{equation}
C_{\phi}^{(k)}(\rho\to\sigma)
\geq
\left[
R_B(\sigma)
-
R_{A\otimes R^{\otimes k}}
\left(
\rho\otimes\phi^{\otimes k}
\right)
\right]_+.
\end{equation}
Moreover,
\begin{equation}
C_{\phi}^{(k)}(\rho\to\sigma)
\geq
\frac{1}{2}
\left[
D_B(\sigma)
-
D_A(\rho)D_R(\phi)^k
\right]_+.
\end{equation}
\end{theorem}
The proof is given in Appendix~\ref{app:assisted-rom-bound}.
Therefore, whenever
\begin{equation}
D_B(\sigma)
>
D_A(\rho)D_R(\phi)^k,
\end{equation}
one necessarily has
\begin{equation}
C_{\phi}^{(k)}(\rho\to\sigma)>0.
\end{equation}

Thus, every situation in which the standard RoM test concludes that $k$ copies contain an insufficient total amount of magic is also detected by the directed conversion deficiency.
Conceptually,
\begin{equation}
\text{standard RoM lower bound}
\subseteq
\text{directed conversion lower bound}.
\end{equation}
The converse, however, does not hold in general.

Even when the RoM-based necessary condition $D_B(\sigma)
\leq
D_A(\rho)D_R(\phi)^k$
is already satisfied, it may still be the case that $C_{\phi}^{(k)}(\rho\to\sigma)>0.$
In this case, the assisted input contains a sufficient total amount of magic according to the scalar RoM comparison, but this does not imply that its magic is available in a form that can be converted exactly into the target under the allowed free operations.
Consequently,
\begin{equation}
k_{\phi}^{\min}(\rho\to\sigma)
\geq
k_{\phi}^{\mathrm{RoM}}(\rho\to\sigma)
\end{equation}
and this inequality can be strict.

More explicitly, suppose that an integer $K$ satisfies
\begin{equation}
K
\geq
k_{\phi}^{\mathrm{RoM}}(\rho\to\sigma).
\end{equation}
At this point, the RoM test cannot determine whether $K$ copies are sufficient.
However, if $C_{\phi}^{(K)}(\rho\to\sigma)>0$
then
\begin{equation}
k_{\phi}^{\min}(\rho\to\sigma)
\geq
K+1
>
k_{\phi}^{\mathrm{RoM}}(\rho\to\sigma).
\end{equation}
This is the sense in which directed conversion geometry can provide a strictly stronger magic-state lower bound than RoM.

A particularly transparent case occurs when
\begin{equation}
R(\rho)=R(\sigma).
\end{equation}
For $k=0$, the RoM comparison provides only the trivial lower bound $k\geq0.$
By contrast, if $C_{\mathcal F}(\rho\to\sigma)>0$,
then conversion without any additional resource state is impossible, and hence
\begin{equation}
k_{\phi}^{\min}(\rho\to\sigma)\geq1.
\end{equation}
Thus, one obtains the strict separation
\begin{equation}
\text{RoM: } k\geq0
\qquad\text{while}\qquad
\text{directed geometry: } k\geq1.
\end{equation}
In the next section, we demonstrate this separation exactly for mixed single-qubit states.

\section{An Exact Single-Qubit Separation Beyond RoM}

In this section, we demonstrate this distinction exactly in the simplest
setting of a single qubit. We construct a continuous family of strictly mixed
states with equal RoM for which free conversion is possible in one direction
but strictly obstructed in the reverse direction. In particular, we obtain
states $\rho_A$ and $\rho_B$ satisfying
\begin{equation}
R(\rho_A)=R(\rho_B),
\end{equation}
while
\begin{equation}
C_{\mathcal F_1}(\rho_A\to\rho_B)=0,
\qquad
C_{\mathcal F_1}(\rho_B\to\rho_A)>0.
\end{equation}
Because all states considered below are strictly mixed, this asymmetry cannot
be attributed to a distinction between pure and mixed states.

\subsection{Stabilizer-Calibrated Geometry for One Qubit}

Every single-qubit state can be written in Bloch form as
\begin{equation}
\rho(\mathbf r)
=
\frac12
\left(
I+r_xX+r_yY+r_zZ
\right),
\qquad
\|\mathbf r\|_2\leq1.
\end{equation}
For a single qubit, the stabilizer-calibrated geometry takes a particularly
simple form.

\begin{proposition}
The single-qubit stabilizer free-state set is
\begin{equation}
\mathcal F_1
=
\left\{
\rho(\mathbf r)
\;\middle|\;
\|\mathbf r\|_1\leq1
\right\}.
\end{equation}
Thus, $\mathcal F_1$ is the stabilizer octahedron in Bloch space.
For a Hermitian observable
\begin{equation}
A
=
a_0I+\mathbf a\cdot\boldsymbol{\sigma},
\end{equation}
the stabilizer-calibrated observable norm is
\begin{equation}
L_{\mathcal F_1}(A)
=
|a_0|+\|\mathbf a\|_\infty.
\end{equation}
Consequently, for arbitrary single-qubit states,
\begin{equation}
d_{\mathcal F_1}
\bigl(
\rho(\mathbf r),\rho(\mathbf s)
\bigr)
=
\|\mathbf r-\mathbf s\|_1.
\end{equation}
Moreover,
\begin{equation}
R(\rho(\mathbf r))
=
\frac12
\left(
\|\mathbf r\|_1-1
\right)_+.
\end{equation}
\end{proposition}

The proof is given in Appendix~\ref{app:single-qubit-geometry}.
Thus, the stabilizer-calibrated geometry of a single qubit coincides with the
ordinary $\ell_1$ geometry of Bloch space. In particular, single-qubit RoM
depends only on the $\ell_1$-distance from the stabilizer octahedron, whereas
$C_{\mathcal F_1}(\rho\to\sigma)$ depends on the position of the target relative
to the entire input-dependent reachable set.

\subsection{A Continuous Family of Equal-RoM Asymmetric Conversions}

Consider the one-parameter family of Bloch vectors
\begin{equation}
\mathbf r_A(a)
=
(a,a,0),\quad
\mathbf r_B(a)
=
\left(
\frac{2a}{3},
\frac{2a}{3},
\frac{2a}{3}
\right),
\end{equation}
with $\frac12<a<\frac1{\sqrt2}$.
We denote the corresponding states by
\begin{equation}
\rho_A(a):=\rho(\mathbf r_A(a)),
\qquad
\rho_B(a):=\rho(\mathbf r_B(a)).
\end{equation}
The main result of this section is the following.

\begin{theorem}
For every $\frac12<a<\frac1{\sqrt2},$
the states $\rho_A(a)$ and $\rho_B(a)$ are both strictly mixed and satisfy
\begin{equation}
R(\rho_A(a))
=
R(\rho_B(a))
=
a-\frac12.
\end{equation}
Nevertheless, their free-conversion properties are directional:
\begin{equation}
C_{\mathcal F_1}
\bigl(
\rho_A(a)\to\rho_B(a)
\bigr)
=
0,
\end{equation}
whereas
\begin{equation}
C_{\mathcal F_1}
\bigl(
\rho_B(a)\to\rho_A(a)
\bigr)
=
(2a-1)
\frac{
3+\sqrt3-\sqrt6-\sqrt2
}{4}
>0.
\end{equation}
\end{theorem}

The proof is given in Appendix~\ref{app:single-qubit-family}.
The forward conversion has a simple free implementation. Let $U$ be a
single-qubit Clifford unitary satisfying
\begin{equation}
UXU^\dagger=Y,
\qquad
UYU^\dagger=Z,
\qquad
UZU^\dagger=X,
\end{equation}
and define the randomized Clifford channel
\begin{equation}
\Lambda_{\mathrm{cyc}}(\rho)
=
\frac13
\left[
\rho
+
U\rho U^\dagger
+
U^2\rho(U^\dagger)^2
\right].
\end{equation}
Then $\Lambda_{\mathrm{cyc}}\in\mathrm{SPO}_1$
and $\Lambda_{\mathrm{cyc}}
\bigl(
\rho_A(a)
\bigr)
=
\rho_B(a)$.
The reverse value is obtained by a matching primal and dual construction, given
in Appendix~\ref{app:single-qubit-family}. Hence, the directional separation is
not an isolated counterexample but persists throughout an open family of
equal-RoM, strictly mixed magic states.

\subsection{A Simple Rational Instance}

A particularly simple member of the family is obtained by setting $a=\frac35.$
Then
\begin{equation}
\mathbf r_A
=
\left(
\frac35,\frac35,0
\right),
\qquad
\mathbf r_B
=
\left(
\frac25,\frac25,\frac25
\right).
\end{equation}
Let
\begin{equation}
\rho_A:=\rho(\mathbf r_A),
\qquad
\rho_B:=\rho(\mathbf r_B).
\end{equation}

Both states are strictly mixed. In fact,
\begin{equation}
\operatorname{Tr}(\rho_A^2)
=
\frac{43}{50}<1,
\qquad
\operatorname{Tr}(\rho_B^2)
=
\frac{37}{50}<1.
\end{equation}
Their RoM values and state-level stabilizer quasiprobability weights coincide:
\begin{equation}
R(\rho_A)
=
R(\rho_B)
=
\frac1{10},
\end{equation}
and
\begin{equation}
D(\rho_A)
=
D(\rho_B)
=
\frac65.
\end{equation}
Nevertheless,
\begin{equation}
C_{\mathcal F_1}(\rho_A\to\rho_B)
=
0,
\end{equation}
whereas
\begin{equation}
C_{\mathcal F_1}(\rho_B\to\rho_A)
=
\frac{
3+\sqrt3-\sqrt6-\sqrt2
}{
20
}
>0.
\end{equation}
Thus, $R(\rho_A)=R(\rho_B)$
does not imply exact free interconvertibility even between strictly mixed
single-qubit states.

The same example also realizes the strict resource-state improvement discussed
in the previous section. For the reverse conversion,
\begin{equation}
R(\rho_B)=R(\rho_A),
\end{equation}
so the RoM comparison with no additional resource state gives only the trivial
bound $k\geq0.$
By contrast,
\begin{equation}
C_{\mathcal F_1}(\rho_B\to\rho_A)>0
\end{equation}
certifies that exact conversion without an additional magic resource state is
impossible, and therefore
\begin{equation}
k_\phi^{\min}(\rho_B\to\rho_A)\geq1.
\end{equation}
Thus, the single-qubit family above demonstrates that equal RoM does not imply
free interconvertibility, equal state-level quasiprobability weights do not
imply equal conversion-specific bottlenecks, and directed conversion geometry
can yield a nontrivial resource-state lower bound even when the RoM-based bound
is trivial.

\section{Summary and Discussion}

In this work, we developed a geometric framework that extends the robustness of magic (RoM) from single-state resource quantification to directed state conversion. The central idea is to replace the fixed stabilizer free-state set appearing in RoM by the input-dependent set of states freely reachable from a given resource state.

Within this framework, RoM is recovered as the distance from a state to the stabilizer free-state set, whereas the directed conversion deficiency
$C_{\mathcal F}(\rho\to\sigma)$ measures the distance from the target $\sigma$ to the freely reachable set generated from the input $\rho$. Thus, RoM quantifies the amount of magic contained in a state, while $C_{\mathcal F}$ characterizes whether that magic is available in a form convertible into a specified target.

The directed conversion deficiency is nonnegative, vanishes exactly when the target is freely reachable from the input, satisfies a directed triangle inequality, and behaves consistently under free preprocessing and postprocessing. For a free input, it reduces exactly to RoM, and the usual conversion obstruction obtained from RoM monotonicity is automatically contained in the directed geometry.

A central result is that this geometric extension also preserves the stabilizer-quasiprobability structure associated with RoM. After optimal free processing of the input, the minimum signed stabilizer weight required to represent the remaining conversion residual is determined by $C_{\mathcal F}$. The corresponding conditional conversion quasiprobability weight is
\begin{equation}
\Gamma_{\mathrm{conv}}(\rho\to\sigma)
=
1+2C_{\mathcal F}(\rho\to\sigma),
\end{equation}
which is the directed-conversion analogue of the state-level relation
$D=1+2R$. In this sense, $C_{\mathcal F}$ quantifies a conversion-specific sign problem that remains after optimal free processing.

This interpretation should not be identified with the total classical complexity of simulating the full input-output process. In particular, $C_{\mathcal F}$ does not include the cost of simulating a magic input itself, and a general stabilizer-preserving operation need not admit an efficient classical implementation. For general magic inputs, the quasiprobability interpretation is therefore naturally conditional or incremental. Restricting the optimization to an explicitly classically simulable subclass of free operations yields a more direct algorithmic interpretation.

We further extended the framework to resource-assisted conversion, in which additional copies of a fixed magic state are supplied. The resulting assisted conversion deficiency determines whether a specified number of resource states is sufficient for conversion and automatically recovers the standard RoM-based lower bound on magic-state consumption. Importantly, it can remain strictly positive even after the scalar RoM comparison becomes inconclusive, and can therefore provide a strictly stronger resource-state lower bound. 

These distinctions were demonstrated exactly in a single-qubit setting. We constructed a continuous family of strictly mixed states with identical RoM for which free conversion is possible in one direction but strictly obstructed in the reverse direction. Hence, equal RoM does not imply free interconvertibility even among mixed states. The same example also shows that identical state-level quasiprobability weights can correspond to different conversion-specific bottlenecks, and that the directed geometry can yield a nontrivial resource-state lower bound even when the RoM-based bound is trivial.

Taken together, these results show that scalar resource content and directed convertibility encode complementary information. The contribution of the present work is therefore not the general notion of a conversion distance itself, but the extension of the stabilizer-adapted geometry, quasiprobability structure, and resource-cost interpretation specific to RoM from individual states to directed magic-state conversion.

\subsection{Outlook}

An important direction for future work is to apply the framework to algorithm-dependent magic-state requirements. For a fixed quantum algorithm, the most useful resource state need not be the one with the largest RoM. Resource states with comparable scalar magic content may differ substantially in their ability to generate the states required by a particular circuit or subroutine. The resource-assisted conversion framework provides a natural way to compare candidate magic states in terms of the number of copies, preparation cost, and conversion deficiency required to reach an algorithmic target.

This perspective may be particularly relevant in fault-tolerant quantum computation, where different nonstabilizer resource states can have very different distillation and preparation costs. Rather than optimizing a magic-state factory independently of its eventual use, one may seek to optimize the resource state together with its conversion into the states required by a specific algorithm. Developing practical methods to determine which type and amount of magic should be prepared for a given computational task is therefore a natural next step.

A second direction is to use directed conversion geometry as an algorithm-specific diagnostic of quantum advantage. Hybrid quantum--classical algorithms such as quantum-selected configuration interaction (QSCI) \cite{kanno2026quantum,nakagawa2024adapt,sugisaki2025hamiltonian,mikkelsen2025quantum,shirai2025enhancing,reinholdt2025critical,robledo2025chemistry} provide a natural setting for such an investigation. Instead of asking only whether a sampled or prepared state possesses nonzero magic, one may ask how difficult it is to reproduce the state required by the relevant sampling or preparation stage from an efficiently simulable reference.

For QSCI and related sampling-based algorithms, it would therefore be useful to study the relation between state-preparation cost, conversion-specific magic, sampling efficiency, and the final computational accuracy. Such an analysis could help distinguish regimes in which the quantum device prepares states that are effectively reproducible by stabilizer-based classical methods from regimes in which an irreducible conversion-specific resource remains.

The directed conversion deficiency alone does not establish a universal classical-complexity lower bound and therefore cannot, by itself, prove quantum advantage. A concrete algorithmic analysis must specify the relevant class of classically simulable operations and compare the resulting resource-theoretic bounds with the best available classical methods and with the end-to-end computational cost. Nevertheless, the directed formulation provides a more task-oriented question: not merely whether magic is present, but whether the particular form of magic required by the computation can be efficiently reproduced from an available classical or stabilizer reference.

Finally, extending the framework to larger systems will require efficient methods for evaluating or bounding reachable-set distances. Semidefinite relaxations, symmetry reduction, tensor-product structure, and optimized witness constructions are promising directions. Such developments would enable applications to concrete multiqubit circuits, magic-state preparation protocols, and quantum algorithms, and may clarify whether conversion-specific magic can serve as a practical resource diagnostic for the design and assessment of quantum computation.

\appendix

\section{Noise Stability and Error Propagation}
\label{app:noise_stability}

In the main text, we established the exact identity
\begin{equation}
    \operatorname{dist}_F(\rho,F)
    =
    2R(\rho),
    \label{eq:appA_dist_R}
\end{equation}
relating the distance from a state to the stabilizer free-state set $F$ in
the stabilizer-calibrated metric $d_F$ to the robustness of magic (RoM).

An immediate consequence of this identity is that the sensitivity of RoM to
noise can be controlled directly by the same resource-adapted metric. In this
appendix, we summarize this stability relation and its implications for error
propagation in multistage processes. We also discuss how the dual-observable
representation of $d_F$ can be used to identify resource-relevant directions
of noise.

\subsection{Metric Stability of RoM under Noise}
\label{app:metric_stability_rom}

Let $\rho$ denote an ideal target state and let $\rho'$ denote the noisy state
obtained experimentally. We model the noise process by a general CPTP map
$\mathcal{N}$,
\begin{equation}
    \rho'
    =
    \mathcal{N}(\rho).
    \label{eq:noisy_state}
\end{equation}
No assumption is made that $\mathcal{N}$ is a free operation. The following
stability bound therefore applies independently of the particular noise
model.

\begin{proposition}[Metric stability of RoM]
\label{prop:appA_rom_stability}
For arbitrary quantum states $\rho$ and $\omega$,
\begin{equation}
        \left|
            R(\rho)-R(\omega)
        \right|
        \leq
        \frac{1}{2}
        d_F(\rho,\omega).
    \label{eq:appA_rom_stability}
\end{equation}
In particular, if $\rho'=\mathcal{N}(\rho)$, then
\begin{equation}
        \left|
            R(\rho)-R(\rho')
        \right|
        \leq
        \frac{1}{2}
        d_F(\rho,\rho').
    \label{eq:appA_noise_stability}
\end{equation}
\end{proposition}

\begin{proof}
The result follows from the general fact that the distance to a set is
$1$-Lipschitz with respect to the underlying metric.

For any $\sigma\in F$, the triangle inequality gives
\begin{equation}
    d_F(\rho,\sigma)
    \leq
    d_F(\rho,\omega)
    +
    d_F(\omega,\sigma).
\end{equation}
Taking the infimum over $\sigma\in F$ yields
\begin{equation}
    \operatorname{dist}_F(\rho,F)
    \leq
    d_F(\rho,\omega)
    +
    \operatorname{dist}_F(\omega,F).
\end{equation}
Hence
\begin{equation}
    \operatorname{dist}_F(\rho,F)
    -
    \operatorname{dist}_F(\omega,F)
    \leq
    d_F(\rho,\omega).
\end{equation}
Interchanging $\rho$ and $\omega$ gives the reverse inequality, and therefore
\begin{equation}
    \left|
        \operatorname{dist}_F(\rho,F)
        -
        \operatorname{dist}_F(\omega,F)
    \right|
    \leq
    d_F(\rho,\omega).
\end{equation}
Using Eq.~\eqref{eq:appA_dist_R} proves
Eq.~\eqref{eq:appA_rom_stability}.
\end{proof}

A useful feature of Eq.~\eqref{eq:appA_noise_stability} is that no additional
geometric constant is required to control the noise-induced variation of
RoM. Since the metric itself is calibrated by the stabilizer free-state set,
\begin{equation}
    \frac{1}{2}d_F(\rho,\rho')
\end{equation}
provides a natural resource-adapted error scale.

For example, if
\begin{equation}
    d_F(\rho,\rho')
    \leq
    \epsilon,
\end{equation}
then
\begin{equation}
    \left|
        R(\rho)-R(\rho')
    \right|
    \leq
    \frac{\epsilon}{2}.
    \label{eq:appA_epsilon_stability}
\end{equation}
This gives a model-independent certification of the discrepancy between the
RoM of an ideal state and that of a noisy experimental state.

\subsection{Error Propagation in Multistage Processes}
\label{app:multistage_error_propagation}

In realistic quantum experiments and fault-tolerant protocols, noise need not
act as a single process. Instead, a state may pass through several stages,
such as state preparation, gate operations, storage, measurements, and
feedback, producing a sequence
\begin{equation}
    \rho_0
    \longrightarrow
    \rho_1
    \longrightarrow
    \cdots
    \longrightarrow
    \rho_m.
    \label{eq:appA_multistage_sequence}
\end{equation}

Define the discrepancy at stage $i$ by
\begin{equation}
    \epsilon_i
    :=
    d_F(\rho_i,\rho_{i+1}).
    \label{eq:appA_stage_error}
\end{equation}
The triangle inequality gives
\begin{equation}
    d_F(\rho_0,\rho_m)
    \leq
    \sum_{i=0}^{m-1}
    d_F(\rho_i,\rho_{i+1}).
    \label{eq:appA_multistage_metric_bound}
\end{equation}
Combining this with Proposition~\ref{prop:appA_rom_stability}, we obtain
\begin{equation}
        \left|
            R(\rho_m)-R(\rho_0)
        \right|
        \leq
        \frac{1}{2}
        \sum_{i=0}^{m-1}
        d_F(\rho_i,\rho_{i+1}).
    \label{eq:appA_multistage_rom_bound}
\end{equation}

More generally, for any intermediate indices $i<j$,
\begin{equation}
    \left|
        R(\rho_j)-R(\rho_i)
    \right|
    \leq
    \frac{1}{2}
    \sum_{\ell=i}^{j-1}
    d_F(\rho_\ell,\rho_{\ell+1}).
    \label{eq:appA_intermediate_bound}
\end{equation}

Thus, if the stabilizer-calibrated discrepancy can be estimated separately at
each stage, the total variation of magic across the entire process can be
bounded additively.

The individual quantities
\begin{equation}
    d_F(\rho_i,\rho_{i+1})
\end{equation}
also provide a natural way to compare which stages contribute most strongly
to the total resource-relevant deviation.

The bound in Eq.~\eqref{eq:appA_multistage_rom_bound} is, however, an
additive worst-case bound. It does not exploit possible cancellation between
error directions generated at different stages. It should therefore be
interpreted as a robust process-level bound obtained from local
resource-relevant discrepancies rather than as a complete description of the
underlying error dynamics.

\subsection{Directional Diagnosis of Magic-Relevant Errors}
\label{app:error_direction_diagnosis}

The stability bound controls the variation of RoM as a scalar quantity, but
the stabilizer-calibrated metric also contains directional information.

Let
\begin{equation}
    \Delta
    :=
    \rho-\rho'
    \label{eq:appA_noise_deviation}
\end{equation}
denote the noise-induced deviation. The dual representation of the metric is
\begin{equation}
    d_F(\rho,\rho')
    =
    \sup_{\substack{
        A=A^\dagger\\
        L_F(A)\leq1
    }}
    \left|
        \operatorname{Tr}(\Delta A)
    \right|.
    \label{eq:appA_dual_noise_metric}
\end{equation}

An optimal observable $A_\star$, or a near-optimal observable, therefore
identifies a measurement direction along which the ideal and noisy states
are most strongly separated on the stabilizer-calibrated scale. In
particular, if
\begin{equation}
    L_F(A_\star)\leq1
\end{equation}
and
\begin{equation}
    \left|
        \operatorname{Tr}(\Delta A_\star)
    \right|
    \simeq
    d_F(\rho,\rho'),
    \label{eq:appA_near_optimal_witness}
\end{equation}
then $A_\star$ captures a dominant resource-relevant direction of the
noise-induced deviation.

For example, consider the Pauli expansion
\begin{equation}
    A_\star
    =
    \sum_P a_P P.
    \label{eq:appA_pauli_expansion}
\end{equation}
The dominant Pauli coefficients may indicate which correlators contribute
most strongly to the discrepancy between the ideal and experimental states.
Dominant low-weight terms may suggest that local deviations are particularly
important, whereas dominant high-weight terms may indicate a significant role
for correlated errors. Concentration of the coefficients within a particular
family of Pauli operators may similarly reveal a preferred
symmetry-breaking or error-sensitive direction.

In this sense, the stabilizer-calibrated geometry naturally addresses the
diagnostic question which observable directions detect the noise most strongly?
rather than providing only a scalar estimate of resource degradation.

An optimizer $A_\star$ should not, however, be interpreted as uniquely
identifying a microscopic physical noise channel. Distinct microscopic error
mechanisms may generate the same or similar expectation-value deviations.
The observable $A_\star$ is more appropriately viewed as a signature of the
resource-relevant discrepancy induced by the noise.

\subsection{Certification from Individual Observables}
\label{app:single_observable_certification}

Even when the full optimization defining $d_F$ is difficult, every feasible
observable provides a certified lower bound.

Let
\begin{equation}
    A=A^\dagger,
    \qquad
    L_F(A)\leq1.
    \label{eq:appA_feasible_observable}
\end{equation}
Then, directly from the definition of $d_F$,
\begin{equation}
        \left|
            \operatorname{Tr}
            \left[
                (\rho-\rho')A
            \right]
        \right|
        \leq
        d_F(\rho,\rho').
    \label{eq:appA_single_observable_distance_bound}
\end{equation}
Thus, a single experimentally accessible expectation-value discrepancy,
\begin{equation}
    \left|
        \langle A\rangle_\rho
        -
        \langle A\rangle_{\rho'}
    \right|,
\end{equation}
already gives a certified lower bound on the ideal--experimental
stabilizer-calibrated distance.

Similarly, for every free state $\sigma\in F$,
\begin{equation}
    \left|
        \operatorname{Tr}
        \left[
            (\rho-\sigma)A
        \right]
    \right|
    \leq
    d_F(\rho,\sigma).
\end{equation}
Taking the infimum over free states yields
\begin{equation}
        \inf_{\sigma\in F}
        \left|
            \operatorname{Tr}
            \left[
                (\rho-\sigma)A
            \right]
        \right|
        \leq
        \operatorname{dist}_F(\rho,F)
        =
        2R(\rho).
    \label{eq:appA_single_observable_rom_bound}
\end{equation}

Hence even a single feasible observable can provide an experimentally
estimable lower bound on RoM. This observation motivates the
measurement-constrained constructions developed in the next appendix.

\section{Proofs for the Stabilizer-Calibrated Geometry}

\subsection{Metric Properties}
\label{app:metric-properties}

We first establish the nondegeneracy of $L_{\mathcal F}$.
The key observation is that stabilizer states span the full real vector space of Hermitian operators.
For an nonidentity $n$-qubit Pauli operator $P$, define $\Pi_\pm(P) = \frac{I\pm P}{2}$. The normalized states $\omega_\pm(P) := \frac{\Pi_\pm(P)}{2^{n-1}}$ are mixed stabilizer states and satisfy $P = 2^{n-1} \left[ \omega_+(P)-\omega_-(P) \right]$. Thus, every Pauli operator can be expressed as a real linear combination of stabilizer states.
The identity operator is also in the real span of stabilizer states, since
$I=2^n(I/2^n)$ and $I/2^n\in\mathcal F$.
Since the Pauli operators form a real basis of the Hermitian operator space, the stabilizer states span the entire Hermitian operator space.
Suppose now that $L_{\mathcal F}(A)=0$. Then $\operatorname{Tr}(\omega A)=0 \qquad \forall\,\omega\in\mathcal F$. Since the stabilizer states span the Hermitian operator space, this implies $A=0$. Hence, $L_{\mathcal F}$ is nondegenerate and is therefore a norm on the real vector space of Hermitian operators.

We next establish the finiteness of $d_{\mathcal F}$.
In finite dimensions, any two norms on the same finite-dimensional vector space are equivalent.
Hence, there exists a finite constant $\alpha_n>0$ such that $\|A\|_\infty \leq \alpha_n L_{\mathcal F}(A)$ for every Hermitian operator $A$.
Therefore, if $L_{\mathcal F}(A)\leq1$, then $\|A\|_\infty \leq \alpha_n$. By H\"older's inequality,
\begin{align}
\left|
\operatorname{Tr}
\left[
(\rho-\sigma)A
\right]
\right|
&\leq
\|\rho-\sigma\|_1
\|A\|_\infty
\nonumber\\
&\leq
\alpha_n
\|\rho-\sigma\|_1.
\end{align}
Taking the supremum over all admissible $A$ gives $d_{\mathcal F}(\rho,\sigma) \leq \alpha_n \|\rho-\sigma\|_1$, and hence $d_{\mathcal F}$ is finite.

Finally, we prove separation.
Suppose that $\rho\neq\sigma$ and define $X:=\rho-\sigma\neq0$. Since $L_{\mathcal F}$ is a norm, $L_{\mathcal F}(X)>0$. Choose $A := \frac{X}{L_{\mathcal F}(X)}$. Then $L_{\mathcal F}(A)=1$, and
\begin{equation}
\operatorname{Tr}(XA)
=
\frac{\operatorname{Tr}(X^2)}
{L_{\mathcal F}(X)}
>0.
\end{equation}
Therefore, $d_{\mathcal F}(\rho,\sigma)>0$. It follows that $d_{\mathcal F}(\rho,\sigma)=0 \iff \rho=\sigma$, so $d_{\mathcal F}$ is a genuine metric on the finite-dimensional quantum state space.

\subsection{Contractivity under Free Operations}
\label{app:contractivity}

Let $\Lambda$ be a free operation satisfying $\Lambda(\mathcal F)\subseteq\mathcal F$. For every Hermitian observable $A$,
\begin{align}
L_{\mathcal F}(\Lambda^\dagger(A))
&=
\sup_{\omega\in\mathcal F}
\left|
\operatorname{Tr}
\left[
\omega\Lambda^\dagger(A)
\right]
\right|
\nonumber\\
&=
\sup_{\omega\in\mathcal F}
\left|
\operatorname{Tr}
\left[
\Lambda(\omega)A
\right]
\right|.
\end{align}
Since $\Lambda(\omega)\in\mathcal F \qquad \forall\,\omega\in\mathcal F$, we have $L_{\mathcal F}(\Lambda^\dagger(A)) \leq L_{\mathcal F}(A)$. Therefore, if $L_{\mathcal F}(A)\leq1$, then $L_{\mathcal F}(\Lambda^\dagger(A)) \leq1$.
Using the definition of $d_{\mathcal F}$, we obtain
\begin{align}
d_{\mathcal F}
\bigl(
\Lambda(\rho),\Lambda(\sigma)
\bigr)
&=
\sup_{\substack{A=A^\dagger\\
L_{\mathcal F}(A)\leq1}}
\left|
\operatorname{Tr}
\left[
(\rho-\sigma)\Lambda^\dagger(A)
\right]
\right|
\nonumber\\
&\leq
d_{\mathcal F}(\rho,\sigma).
\end{align}
Hence,
\begin{equation}
d_{\mathcal F}
\bigl(
\Lambda(\rho),\Lambda(\sigma)
\bigr)
\leq
d_{\mathcal F}(\rho,\sigma).
\end{equation}
Thus, applying the same free operation to both states cannot increase their stabilizer-calibrated distance.

\subsection{Exact Distance Representation of RoM}
\label{app:rom-distance}

We first prove $2R(\rho) \leq \operatorname{dist}_{\mathcal F} (\rho,\mathcal F)$.
Fix an arbitrary free state $\omega\in\mathcal F$.
Every feasible witness $W$ in the dual formulation of RoM satisfies
\begin{equation}
L_{\mathcal F}(W)
=
\sup_{\tau\in\mathcal F}
|\operatorname{Tr}(\tau W)|
\leq1.
\end{equation}
Since $\omega\in\mathcal F$, we also have $\operatorname{Tr}(\omega W)\leq1$. Therefore,
\begin{align}
\operatorname{Tr}(\rho W)-1
&\leq
\operatorname{Tr}(\rho W)
-
\operatorname{Tr}(\omega W)
\nonumber\\
&=
\operatorname{Tr}
\left[
(\rho-\omega)W
\right]
\nonumber\\
&\leq
\left|
\operatorname{Tr}
\left[
(\rho-\omega)W
\right]
\right|
\nonumber\\
&\leq
d_{\mathcal F}(\rho,\omega).
\end{align}
Taking the supremum over all feasible witnesses $W$ and using the dual representation of RoM gives $2R(\rho) \leq d_{\mathcal F}(\rho,\omega)$. Since $\omega\in\mathcal F$ was arbitrary,
\begin{equation}
2R(\rho)
\leq
\inf_{\omega\in\mathcal F}
d_{\mathcal F}(\rho,\omega)
=
\operatorname{dist}_{\mathcal F}
(\rho,\mathcal F).
\end{equation}

We next prove the reverse inequality, $\operatorname{dist}_{\mathcal F} (\rho,\mathcal F) \leq 2R(\rho)$. Set $t:=R(\rho)$. By the definition of RoM, there exist $\sigma,\tau\in\mathcal F$ such that $\rho = (1+t)\sigma-t\tau$. Equivalently, $\rho-\sigma = t(\sigma-\tau)$. Since $\sigma\in\mathcal F$,
\begin{equation}
\operatorname{dist}_{\mathcal F}
(\rho,\mathcal F)
\leq
d_{\mathcal F}(\rho,\sigma).
\end{equation}
Moreover,
\begin{align}
d_{\mathcal F}(\rho,\sigma)
&=
t
\sup_{\substack{A=A^\dagger\\
L_{\mathcal F}(A)\leq1}}
\left|
\operatorname{Tr}
\left[
(\sigma-\tau)A
\right]
\right|.
\end{align}
Since $L_{\mathcal F}(A)\leq1$ and $\sigma,\tau\in\mathcal F$,
\begin{equation}
|\operatorname{Tr}(\sigma A)|
\leq1,
\qquad
|\operatorname{Tr}(\tau A)|
\leq1.
\end{equation}
Hence,
\begin{align}
\left|
\operatorname{Tr}
\left[
(\sigma-\tau)A
\right]
\right|
&\leq
|\operatorname{Tr}(\sigma A)|
+
|\operatorname{Tr}(\tau A)|
\nonumber\\
&\leq2.
\end{align}
Therefore, $d_{\mathcal F}(\rho,\sigma) \leq 2t = 2R(\rho)$, and consequently $\operatorname{dist}_{\mathcal F} (\rho,\mathcal F) \leq 2R(\rho)$.

Combining the two inequalities, we obtain $\operatorname{dist}_{\mathcal F} (\rho,\mathcal F) = 2R(\rho)$. Equivalently,
\begin{equation}
R(\rho)
=
\frac{1}{2}
\operatorname{dist}_{\mathcal F}
(\rho,\mathcal F).
\end{equation}

\subsection{Stability of RoM}
\label{app:rom-stability}

For an arbitrary metric $d$ and a nonempty subset $S$, the distance-to-set function is $1$-Lipschitz:
\begin{equation}
\left|
\operatorname{dist}_{d}(x,S)
-
\operatorname{dist}_{d}(y,S)
\right|
\leq
d(x,y).
\end{equation}
Indeed, for every $s\in S$, the triangle inequality gives $d(x,s) \leq d(x,y)+d(y,s)$. Taking the infimum over $s\in S$ yields
\begin{equation}
\operatorname{dist}_{d}(x,S)
\leq
d(x,y)
+
\operatorname{dist}_{d}(y,S).
\end{equation}
Therefore,
\begin{equation}
\operatorname{dist}_{d}(x,S)
-
\operatorname{dist}_{d}(y,S)
\leq
d(x,y).
\end{equation}
Interchanging $x$ and $y$ gives the reverse inequality, and hence
\begin{equation}
\left|
\operatorname{dist}_{d}(x,S)
-
\operatorname{dist}_{d}(y,S)
\right|
\leq
d(x,y).
\end{equation}
Applying this result to $d_{\mathcal F}$ and $\mathcal F$, we obtain
\begin{equation}
\left|
\operatorname{dist}_{\mathcal F}
(\rho,\mathcal F)
-
\operatorname{dist}_{\mathcal F}
(\sigma,\mathcal F)
\right|
\leq
d_{\mathcal F}(\rho,\sigma).
\end{equation}
Using $\operatorname{dist}_{\mathcal F} (\rho,\mathcal F) = 2R(\rho)$, we finally obtain $|R(\rho)-R(\sigma)| \leq \frac{1}{2} d_{\mathcal F}(\rho,\sigma)$.

\section{Proofs for the Directed Conversion Geometry}

\subsection{Zero Set and Exact Convertibility}
\label{app:zero-conversion}

First, suppose that there exists $\Lambda\in\mathrm{SPO}$ such that $\Lambda(\rho)=\sigma$. Then $\sigma\in\mathcal O(\rho)$, and therefore $\operatorname{dist}_{\mathcal F} (\sigma,\mathcal O(\rho)) = 0$. Hence, $C_{\mathcal F}(\rho\to\sigma)=0$.

Conversely, suppose that $C_{\mathcal F}(\rho\to\sigma)=0$.
In finite dimensions, the set of CPTP maps is compact, and the condition $\Lambda(\mathcal F)\subseteq\mathcal F$ is closed. Hence, $\mathrm{SPO}$ is compact.
Moreover, the map $\Lambda\longmapsto\Lambda(\rho)$ is continuous, so its image $\mathcal O(\rho)$ is also compact and therefore closed.
Since $d_{\mathcal F}$ is a genuine metric, $\operatorname{dist}_{\mathcal F} (\sigma,\mathcal O(\rho)) = 0$ implies $\sigma\in\mathcal O(\rho)$. Therefore, there exists some $\Lambda\in\mathrm{SPO}$ such that $\Lambda(\rho)=\sigma$.
Hence,
\begin{equation}
C_{\mathcal F}(\rho\to\sigma)=0
\iff
\exists\,\Lambda\in\mathrm{SPO}
\text{ such that }
\Lambda(\rho)=\sigma.
\end{equation}

\subsection{Basic Properties}
\label{app:conversion-properties}

Non-negativity, $C_{\mathcal F}(\rho\to\sigma)\geq0$, follows immediately from the definition. Since the identity channel belongs to $\mathrm{SPO}$, $C_{\mathcal F}(\rho\to\rho)=0$.

\subsubsection{Directed Triangle Inequality}

For arbitrary $\epsilon>0$, choose $\Lambda,\Gamma\in\mathrm{SPO}$ such that
\begin{equation}
\frac{1}{2}
d_{\mathcal F}
\bigl(
\Lambda(\rho),\sigma
\bigr)
\leq
C_{\mathcal F}(\rho\to\sigma)+\epsilon
\end{equation}
and
\begin{equation}
\frac{1}{2}
d_{\mathcal F}
\bigl(
\Gamma(\sigma),\tau
\bigr)
\leq
C_{\mathcal F}(\sigma\to\tau)+\epsilon.
\end{equation}
Since $\mathrm{SPO}$ is closed under composition, $\Gamma\circ\Lambda\in\mathrm{SPO}$. Therefore,
\begin{equation}
C_{\mathcal F}(\rho\to\tau)
\leq
\frac{1}{2}
d_{\mathcal F}
\bigl(
\Gamma(\Lambda(\rho)),\tau
\bigr).
\end{equation}
Using the triangle inequality for $d_{\mathcal F}$,
\begin{align}
d_{\mathcal F}
\bigl(
\Gamma(\Lambda(\rho)),\tau
\bigr)
&\leq
d_{\mathcal F}
\bigl(
\Gamma(\Lambda(\rho)),\Gamma(\sigma)
\bigr)
\nonumber\\
&\quad+
d_{\mathcal F}
\bigl(
\Gamma(\sigma),\tau
\bigr).
\end{align}
Furthermore, contractivity of $d_{\mathcal F}$ under free operations gives
\begin{equation}
d_{\mathcal F}
\bigl(
\Gamma(\Lambda(\rho)),\Gamma(\sigma)
\bigr)
\leq
d_{\mathcal F}
\bigl(
\Lambda(\rho),\sigma
\bigr).
\end{equation}
Hence,
\begin{equation}
C_{\mathcal F}(\rho\to\tau)
\leq
C_{\mathcal F}(\rho\to\sigma)
+
C_{\mathcal F}(\sigma\to\tau)
+
2\epsilon.
\end{equation}
Since $\epsilon>0$ is arbitrary,
\begin{equation}
C_{\mathcal F}(\rho\to\tau)
\leq
C_{\mathcal F}(\rho\to\sigma)
+
C_{\mathcal F}(\sigma\to\tau).
\end{equation}

\subsubsection{Free Preprocessing of the Input}

Let $\Gamma\in\mathrm{SPO}$. For any $\Lambda\in\mathrm{SPO}$, $\Lambda(\Gamma(\rho)) = (\Lambda\circ\Gamma)(\rho)$, and $\Lambda\circ\Gamma\in\mathrm{SPO}$. Therefore, $\mathcal O(\Gamma(\rho)) \subseteq \mathcal O(\rho)$.
Since the distance from the target to a smaller reachable set cannot be smaller,
\begin{equation}
C_{\mathcal F}
\bigl(
\Gamma(\rho)\to\sigma
\bigr)
\geq
C_{\mathcal F}(\rho\to\sigma).
\end{equation}

\subsubsection{Free Postprocessing of the Target}

For any $\omega\in\mathcal O(\rho)$, there exists $\Lambda\in\mathrm{SPO}$ such that $\omega=\Lambda(\rho)$. Hence, $\Gamma(\omega) = (\Gamma\circ\Lambda)(\rho) \in \mathcal O(\rho)$.
Moreover, by contractivity of $d_{\mathcal F}$,
\begin{equation}
d_{\mathcal F}
\bigl(
\Gamma(\omega),\Gamma(\sigma)
\bigr)
\leq
d_{\mathcal F}(\omega,\sigma).
\end{equation}
Taking the infimum over $\omega\in\mathcal O(\rho)$ gives
\begin{equation}
C_{\mathcal F}
\bigl(
\rho\to\Gamma(\sigma)
\bigr)
\leq
C_{\mathcal F}(\rho\to\sigma).
\end{equation}

\subsection{Bounds in Terms of RoM}
\label{app:conversion-rom-bounds}

We first prove the lower bound.
For arbitrary $\Lambda\in\mathrm{SPO}$, the stability relation for RoM gives
\begin{equation}
|R(\sigma)-R(\Lambda(\rho))|
\leq
\frac{1}{2}
d_{\mathcal F}
\bigl(
\sigma,\Lambda(\rho)
\bigr).
\end{equation}
Therefore,
\begin{equation}
R(\sigma)-R(\Lambda(\rho))
\leq
\frac{1}{2}
d_{\mathcal F}
\bigl(
\sigma,\Lambda(\rho)
\bigr).
\end{equation}
By monotonicity of RoM, $R(\Lambda(\rho)) \leq R(\rho)$. Hence,
\begin{equation}
R(\sigma)-R(\rho)
\leq
\frac{1}{2}
d_{\mathcal F}
\bigl(
\sigma,\Lambda(\rho)
\bigr).
\end{equation}
Taking the infimum over $\Lambda\in\mathrm{SPO}$ yields $R(\sigma)-R(\rho) \leq C_{\mathcal F}(\rho\to\sigma)$. Since $C_{\mathcal F}\geq0$, we obtain $[R(\sigma)-R(\rho)]_+ \leq C_{\mathcal F}(\rho\to\sigma)$.

We next prove the upper bound.
Any free state $\omega\in\mathcal F$ can be generated from an arbitrary input state $\rho$ by a replacement channel that discards the input and prepares $\omega$. Therefore, $\mathcal F \subseteq \mathcal O(\rho)$.
Since the distance to a larger set cannot exceed the distance to a smaller set,
\begin{equation}
\operatorname{dist}_{\mathcal F}
\bigl(
\sigma,\mathcal O(\rho)
\bigr)
\leq
\operatorname{dist}_{\mathcal F}
(\sigma,\mathcal F).
\end{equation}
Therefore, $C_{\mathcal F}(\rho\to\sigma) \leq R(\sigma)$.

Combining the two inequalities gives
\begin{equation}
[R(\sigma)-R(\rho)]_+
\leq
C_{\mathcal F}(\rho\to\sigma)
\leq
R(\sigma).
\end{equation}

\section{Proofs for the Conversion-Simulation Interpretation}

\subsection{Signed Stabilizer Norm}
\label{app:residual-norm}

Let $\{\phi_i\}_{i=1}^{N_{\mathcal F}}$ denote the finite set of pure stabilizer projectors. Since stabilizer projectors span the full Hermitian operator space, every Hermitian operator $X$ admits a decomposition $X=\sum_iq_i\phi_i$.
Consider the optimization problem $\inf \left\{ \sum_i|q_i| \;\middle|\; X=\sum_iq_i\phi_i \right\}$.
Writing $q_i=u_i-v_i, \qquad u_i,v_i\geq0$, turns this into a finite-dimensional linear program. Its dual problem is
\begin{equation}
\sup
\left\{
\operatorname{Tr}(XA)
\;\middle|\;
A=A^\dagger,\;
|\operatorname{Tr}(\phi_iA)|\leq1
\quad
\forall i
\right\}.
\end{equation}
Since $\mathcal F$ is the convex hull of pure stabilizer states, these constraints are equivalent to
\begin{equation}
L_{\mathcal F}(A)
=
\sup_{\omega\in\mathcal F}
|\operatorname{Tr}(\omega A)|
\leq1.
\end{equation}
Moreover, since the feasible set is symmetric under $A\mapsto-A$, introducing an absolute value in the objective does not change the optimal value.
Linear-programming duality therefore gives
\begin{equation}
\|X\|_{\mathcal F}
=
\inf
\left\{
\sum_j|q_j|
\;\middle|\;
X=\sum_jq_j\omega_j,\quad
\omega_j\in\mathcal F
\right\}.
\end{equation}
Now assume $\operatorname{Tr}(X)=0$. For any decomposition $X=\sum_jq_j\omega_j$, each $\omega_j$ has unit trace, and hence $\sum_jq_j=0$.
Let $t_+$ denote the total positive coefficient weight and $t_-$ the total magnitude of the negative coefficient weight. Then $t_+=t_-$. Writing their common value as $t$, and grouping the positive and negative parts into normalized free states, we obtain $X=t(\omega_+-\omega_-), \qquad \omega_\pm\in\mathcal F$.
The total absolute weight of this decomposition is $2t$. Therefore,
\begin{equation}
\|X\|_{\mathcal F}
=
2
\inf
\left\{
t\geq0
\;\middle|\;
X=t(\omega_+-\omega_-),\quad
\omega_\pm\in\mathcal F
\right\}.
\end{equation}

\subsection{Conversion Quasiprobability Weight}
\label{app:conversion-weight}

Fix $\Lambda\in\mathrm{SPO}$ and set $X = \sigma-\Lambda(\rho)$.
Since $\operatorname{Tr}X=0$, the result of the preceding subsection gives
\begin{equation}
\inf
\left\{
\sum_j|q_j|
\;\middle|\;
\sigma-\Lambda(\rho)
=
\sum_jq_j\omega_j
\right\}
=
\|X\|_{\mathcal F}.
\end{equation}
Furthermore,
\begin{equation}
\|X\|_{\mathcal F}
=
d_{\mathcal F}
\bigl(
\Lambda(\rho),\sigma
\bigr).
\end{equation}
Hence,
\begin{equation}
\inf
\left\{
\sum_j|q_j|
\;\middle|\;
\sigma-\Lambda(\rho)
=
\sum_jq_j\omega_j
\right\}
=
d_{\mathcal F}
\bigl(
\Lambda(\rho),\sigma
\bigr).
\end{equation}
Taking the infimum over $\Lambda\in\mathrm{SPO}$ gives
\begin{equation}
\inf_{\Lambda\in\mathrm{SPO}}
d_{\mathcal F}
\bigl(
\Lambda(\rho),\sigma
\bigr)
=
2C_{\mathcal F}(\rho\to\sigma).
\end{equation}
Adding the fixed coefficient $1$ corresponding to the reference branch yields
\begin{equation}
\Gamma_{\mathrm{conv}}(\rho\to\sigma)
=
1+2C_{\mathcal F}(\rho\to\sigma).
\end{equation}

\subsection{Monte Carlo Estimation of the Conversion Residual}
\label{app:residual-monte-carlo}

Let the conversion residual be decomposed as $\sigma-\Lambda(\rho) = \sum_jq_j\omega_j$, and define $\gamma_{\mathrm{res}} = \sum_j|q_j|$.
For a Pauli observable $P$, we wish to estimate
\begin{equation}
\Delta_P(\Lambda)
=
\operatorname{Tr}
\left[
P
\left(
\sigma-\Lambda(\rho)
\right)
\right].
\end{equation}
In each Monte Carlo trial, choose $j$ according to $p_j = \frac{|q_j|}{\gamma_{\mathrm{res}}}$, and sample a Pauli measurement outcome $m\in\{-1,+1\}$ from the stabilizer state $\omega_j$.
Define the random variable $Z = \gamma_{\mathrm{res}} \operatorname{sgn}(q_j)m$. Then
\begin{align}
\mathbb E[Z]
&=
\sum_jq_j
\operatorname{Tr}(P\omega_j)
\nonumber\\
&=
\operatorname{Tr}
\left[
P
\left(
\sigma-\Lambda(\rho)
\right)
\right]
\nonumber\\
&=
\Delta_P(\Lambda).
\end{align}
Thus, $Z$ is an unbiased estimator.
Moreover, $|Z|\leq\gamma_{\mathrm{res}}$. Hoeffding's inequality therefore implies that
\begin{equation}
N_{\mathrm{res}}
\geq
\frac{
2\gamma_{\mathrm{res}}^2
}{
\epsilon^2
}
\log
\left(
\frac{2}{\delta}
\right)
\end{equation}
samples are sufficient to achieve additive error $\epsilon$ with failure probability at most $\delta$.
Optimizing over both the free operation and the residual decomposition gives $\gamma_{\mathrm{res}}^{\min} = 2C_{\mathcal F}(\rho\to\sigma)$, and hence
\begin{equation}
N_{\mathrm{res}}
\geq
\frac{
8C_{\mathcal F}(\rho\to\sigma)^2
}{
\epsilon^2
}
\log
\left(
\frac{2}{\delta}
\right).
\end{equation}

\subsection{Conditional Reconstruction of the Target}
\label{app:conditional-reconstruction}

Consider the decomposition $\sigma = \Lambda(\rho) + \sum_{j\geq1}q_j\omega_j$. Define $c_0:=1, \qquad \eta_0:=\Lambda(\rho)$, and, for $j\geq1$, $c_j:=q_j, \qquad \eta_j:=\omega_j$.
The total absolute weight is $\Gamma := \sum_{j\geq0}|c_j| = 1+\sum_{j\geq1}|q_j|$.
In each trial, choose branch $j$ with probability $\frac{|c_j|}{\Gamma}$, sample a Pauli measurement outcome $m\in\{-1,+1\}$ from the state $\eta_j$, and define $W = \Gamma \operatorname{sgn}(c_j)m$.
Then
\begin{align}
\mathbb E[W]
&=
\sum_jc_j\operatorname{Tr}(P\eta_j)
\nonumber\\
&=
\operatorname{Tr}(P\sigma).
\end{align}
Moreover, $|W|\leq\Gamma$.
Hoeffding's inequality therefore gives the sufficient sample bound
\begin{equation}
N_{\mathrm{cond}}
\geq
\frac{
2\Gamma^2
}{
\epsilon^2
}
\log
\left(
\frac{2}{\delta}
\right).
\end{equation}
After optimization,
\begin{equation}
\Gamma
=
\Gamma_{\mathrm{conv}}(\rho\to\sigma)
=
1+2C_{\mathcal F}(\rho\to\sigma),
\end{equation}
and hence
\begin{equation}
N_{\mathrm{cond}}
\geq
\frac{
2
\left(
1+2C_{\mathcal F}(\rho\to\sigma)
\right)^2
}{
\epsilon^2
}
\log
\left(
\frac{2}{\delta}
\right).
\end{equation}

\subsection{Fully Classical Conversion-Respecting Simulation}
\label{app:full-path-simulation}

Let an optimal stabilizer quasiprobability decomposition of the input state be
\begin{equation}
\rho
=
\sum_i x_i\phi_i,
\qquad
\sum_i |x_i|
=
D(\rho).
\end{equation}
Suppose that
$\Lambda\in\mathrm{SPO}$
admits efficient classical sampling on each stabilizer input $\phi_i$.
For this fixed $\Lambda$, write
\begin{equation}
\sigma
=
\Lambda(\rho)
+
\sum_j q_j\omega_j,
\qquad
\omega_j\in\mathcal F.
\end{equation}
By linearity,
\begin{equation}
\sigma
=
\sum_i x_i\Lambda(\phi_i)
+
\sum_j q_j\omega_j.
\end{equation}

Since $\Lambda$ is classically simulable on each stabilizer input $\phi_i$,
the first term can be sampled branch by branch.
For fixed $\Lambda$, Proposition~\ref{prop:signed-stabilizer-norm} gives
\begin{equation}
\inf_{\{q_j,\omega_j\}}
\sum_j |q_j|
=
d_{\mathcal F}
\bigl(
\Lambda(\rho),\sigma
\bigr).
\end{equation}
Therefore, choosing an optimal residual decomposition for this $\Lambda$,
the total absolute weight of the resulting quasiprobability representation
is bounded by
\begin{equation}
D(\rho)
+
d_{\mathcal F}
\bigl(
\Lambda(\rho),\sigma
\bigr).
\end{equation}

Applying the same Monte Carlo construction as above, together with
Hoeffding's inequality, gives the sufficient sample bound
\begin{equation}
N_{\mathrm{path}}
\geq
\frac{
2
\left[
D(\rho)
+
d_{\mathcal F}
\bigl(
\Lambda(\rho),\sigma
\bigr)
\right]^2
}{
\epsilon^2
}
\log
\left(
\frac{2}{\delta}
\right).
\end{equation}

In particular, suppose that there exists a classically simulable
$\Lambda_\star\in\mathrm{SPO}$ attaining the optimization defining
$C_{\mathcal F}(\rho\to\sigma)$.
Then
\begin{equation}
d_{\mathcal F}
\bigl(
\Lambda_\star(\rho),\sigma
\bigr)
=
2C_{\mathcal F}(\rho\to\sigma),
\end{equation}
and hence the total absolute weight is bounded by
\begin{equation}
D(\rho)
+
2C_{\mathcal F}(\rho\to\sigma).
\end{equation}
Consequently,
\begin{equation}
N_{\mathrm{path}}
\geq
\frac{
2
\left[
D(\rho)
+
2C_{\mathcal F}(\rho\to\sigma)
\right]^2
}{
\epsilon^2
}
\log
\left(
\frac{2}{\delta}
\right)
\end{equation}
samples are sufficient.

Using
\begin{equation}
D(\rho)=1+2R(\rho),
\end{equation}
this becomes
\begin{equation}
N_{\mathrm{path}}
\geq
\frac{
2
\left[
1+2R(\rho)
+
2C_{\mathcal F}(\rho\to\sigma)
\right]^2
}{
\epsilon^2
}
\log
\left(
\frac{2}{\delta}
\right).
\end{equation}

This is a constructive upper bound for the particular
conversion-respecting simulation strategy considered here.
If additional cancellations occur between the state-level and residual
signed contributions, the actual total variation may be smaller.

\subsection{Restriction to Classically Simulable Free Operations}
\label{app:spo-sim}

Let $\mathrm{SPO}_{\mathrm{sim}} \subseteq \mathrm{SPO}$ denote a subclass of free operations that admit efficient classical sampling on stabilizer inputs.
We define
\begin{equation}
C_{\mathcal F}^{\mathrm{sim}}
(\rho\to\sigma)
=
\frac{1}{2}
\inf_{\Lambda\in\mathrm{SPO}_{\mathrm{sim}}}
d_{\mathcal F}
\bigl(
\Lambda(\rho),\sigma
\bigr).
\end{equation}
The same argument as above yields
\begin{equation}
\Gamma_{\mathrm{conv}}^{\mathrm{sim}}
(\rho\to\sigma)
=
1+
2C_{\mathcal F}^{\mathrm{sim}}
(\rho\to\sigma).
\end{equation}
Since $\mathrm{SPO}_{\mathrm{sim}} \subseteq \mathrm{SPO}$, the optimization is performed over a smaller set, and therefore
\begin{equation}
C_{\mathcal F}(\rho\to\sigma)
\leq
C_{\mathcal F}^{\mathrm{sim}}(\rho\to\sigma).
\end{equation}
Thus, optimization over the full SPO class yields the information-theoretic minimum signed correction, whereas restriction to $\mathrm{SPO}_{\mathrm{sim}}$ yields the quantity corresponding to a literal algorithmic classical-simulation interpretation.

\section{Proofs for Resource-Assisted Conversion Bounds}

\subsection{Resource-Assisted Conversion and Exact Cost}
\label{app:assisted-conversion}

We first prove the equivalence between $C_{\phi}^{(k)}(\rho\to\sigma)=0$ and exact convertibility.
Suppose that there exists $\Lambda \in \mathrm{SPO} (A\otimes R^{\otimes k}\to B)$ such that $\Lambda \left( \rho\otimes\phi^{\otimes k} \right) = \sigma$. Then, by definition, $C_{\phi}^{(k)}(\rho\to\sigma)=0$.

Conversely, suppose that $C_{\phi}^{(k)}(\rho\to\sigma)=0$. In finite dimensions, the corresponding set of SPO maps is compact, and the map $\Lambda \longmapsto \Lambda \left( \rho\otimes\phi^{\otimes k} \right)$ is continuous. Hence, the reachable set generated from the assisted input is compact and therefore closed.
Since $d_{\mathcal F_B}$ is a genuine metric, zero distance from the target to this reachable set implies that the target itself belongs to the reachable set. Therefore, there exists a free operation realizing the exact conversion.

We next prove $C_{\phi}^{(k+1)}(\rho\to\sigma) \leq C_{\phi}^{(k)}(\rho\to\sigma)$.
If $k+1$ copies of $\phi$ are available, one may discard one copy and implement any protocol available with only $k$ copies. Therefore, the reachable set generated using $k+1$ copies contains the reachable set generated using $k$ copies.
Since the distance to a larger set cannot exceed the distance to a smaller set, $C_{\phi}^{(k+1)}(\rho\to\sigma) \leq C_{\phi}^{(k)}(\rho\to\sigma)$.

Finally, if $C_{\phi}^{(K)}(\rho\to\sigma)>0$, then exact conversion is impossible using $K$ copies. By the definition of $k_{\phi}^{\min}$, $k_{\phi}^{\min}(\rho\to\sigma) \geq K+1$.

\subsection{Recovery of the RoM Lower Bound}
\label{app:assisted-rom-bound}

Set $\eta_k := \rho\otimes\phi^{\otimes k}$.
For every $\Lambda \in \mathrm{SPO} (A\otimes R^{\otimes k}\to B)$, the stability relation for RoM gives
\begin{equation}
\left|
R_B(\sigma)
-
R_B(\Lambda(\eta_k))
\right|
\leq
\frac{1}{2}
d_{\mathcal F_B}
\bigl(
\Lambda(\eta_k),\sigma
\bigr).
\end{equation}
Therefore,
\begin{equation}
\frac{1}{2}
d_{\mathcal F_B}
\bigl(
\Lambda(\eta_k),\sigma
\bigr)
\geq
R_B(\sigma)
-
R_B(\Lambda(\eta_k)).
\end{equation}
By monotonicity of RoM, $R_B(\Lambda(\eta_k)) \leq R_{A\otimes R^{\otimes k}}(\eta_k)$, and hence
\begin{equation}
\frac{1}{2}
d_{\mathcal F_B}
\bigl(
\Lambda(\eta_k),\sigma
\bigr)
\geq
\left[
R_B(\sigma)
-
R_{A\otimes R^{\otimes k}}(\eta_k)
\right]_+.
\end{equation}
Taking the infimum over all $\Lambda$ gives
\begin{equation}
C_{\phi}^{(k)}(\rho\to\sigma)
\geq
\left[
R_B(\sigma)
-
R_{A\otimes R^{\otimes k}}(\eta_k)
\right]_+.
\end{equation}
Next, using $D_X=1+2R_X$ and $D_{A\otimes R^{\otimes k}}(\eta_k) \leq D_A(\rho)D_R(\phi)^k$, the preceding inequality yields
\begin{equation}
C_{\phi}^{(k)}(\rho\to\sigma)
\geq
\frac{1}{2}
\left[
D_B(\sigma)
-
D_{A\otimes R^{\otimes k}}(\eta_k)
\right]_+.
\end{equation}
By submultiplicativity, $D_{A\otimes R^{\otimes k}}(\eta_k) \leq D_A(\rho)D_R(\phi)^k$, and therefore
\begin{equation}
C_{\phi}^{(k)}(\rho\to\sigma)
\geq
\frac{1}{2}
\left[
D_B(\sigma)
-
D_A(\rho)D_R(\phi)^k
\right]_+.
\end{equation}

In particular, if $D_B(\sigma) > D_A(\rho)D_R(\phi)^k$, then $C_{\phi}^{(k)}(\rho\to\sigma)>0$.

Conversely, if exact conversion is possible, then $C_{\phi}^{(k)}(\rho\to\sigma)=0$, so it is necessary that $D_B(\sigma) \leq D_A(\rho)D_R(\phi)^k$.
Assuming $D_R(\phi)>1$, taking logarithms gives $k \geq \frac{ \log D_B(\sigma)-\log D_A(\rho) }{ \log D_R(\phi) }$. Since $k$ is a nonnegative integer, $k \geq k_{\phi}^{\mathrm{RoM}}(\rho\to\sigma)$.

\subsection{Strict Strengthening Beyond RoM}
\label{app:strict-resource-improvement}

Every $k$ for which exact conversion is possible must satisfy the RoM necessary condition. Therefore,
\begin{equation}
k_{\phi}^{\min}(\rho\to\sigma)
\geq
k_{\phi}^{\mathrm{RoM}}(\rho\to\sigma).
\end{equation}

Now suppose that $K \geq k_{\phi}^{\mathrm{RoM}}(\rho\to\sigma)$ but $C_{\phi}^{(K)}(\rho\to\sigma)>0$.
The latter condition means that exact conversion using $K$ copies is impossible. Hence, $k_{\phi}^{\min}(\rho\to\sigma) \geq K+1$.
Since $K \geq k_{\phi}^{\mathrm{RoM}}(\rho\to\sigma)$, we have $K+1 > k_{\phi}^{\mathrm{RoM}}(\rho\to\sigma)$.
Therefore,
\begin{equation}
k_{\phi}^{\min}(\rho\to\sigma)
\geq
K+1
>
k_{\phi}^{\mathrm{RoM}}(\rho\to\sigma).
\end{equation}

In particular, if $R(\rho)=R(\sigma)$ and $C_{\mathcal F}(\rho\to\sigma)>0$, then the RoM test gives only $k\geq0$, whereas the directed conversion deficiency gives $k_{\phi}^{\min}(\rho\to\sigma)\geq1$.

\section{Proofs for the Exact Single-Qubit Separation}

\subsection{Single-Qubit Stabilizer-Calibrated Geometry}
\label{app:single-qubit-geometry}

Every single-qubit state has the Bloch representation
\begin{equation}
\rho(\mathbf r)
=
\frac12
\left(
I+\mathbf r\cdot\boldsymbol{\sigma}
\right),
\qquad
\|\mathbf r\|_2\leq1.
\end{equation}
The pure single-qubit stabilizer states are the six Pauli eigenstates, whose
Bloch vectors are $\pm\mathbf e_x, \qquad \pm\mathbf e_y, \qquad \pm\mathbf e_z$. Their convex hull is the $\ell_1$-unit ball. Therefore,
\begin{equation}
\mathcal F_1
=
\left\{
\rho(\mathbf r)
\;\middle|\;
\|\mathbf r\|_1\leq1
\right\}.
\end{equation}
Now let $A = a_0I+\mathbf a\cdot\boldsymbol{\sigma}$ be Hermitian. For $\rho(\mathbf t)\in\mathcal F_1$, we have
\begin{equation}
\operatorname{Tr}
\left[
\rho(\mathbf t)A
\right]
=
a_0+\mathbf a\cdot\mathbf t.
\end{equation}
Hence,
\begin{equation}
L_{\mathcal F_1}(A)
=
\sup_{\|\mathbf t\|_1\leq1}
\left|
a_0+\mathbf a\cdot\mathbf t
\right|.
\end{equation}
By $\ell_1$--$\ell_\infty$ duality,
\begin{equation}
\sup_{\|\mathbf t\|_1\leq1}
|\mathbf a\cdot\mathbf t|
=
\|\mathbf a\|_\infty,
\end{equation}
and the extremal sign can be chosen to align with $a_0$. Therefore, $L_{\mathcal F_1}(A) = |a_0|+\|\mathbf a\|_\infty$.
For two states $\rho(\mathbf r)$ and $\rho(\mathbf s)$,
\begin{equation}
\operatorname{Tr}
\left[
(\rho(\mathbf r)-\rho(\mathbf s))A
\right]
=
(\mathbf r-\mathbf s)\cdot\mathbf a.
\end{equation}
The scalar component $a_0$ does not contribute. Thus, in optimizing the
distance, one may take $a_0=0$, and the constraint $L_{\mathcal F_1}(A)\leq1$ reduces to $\|\mathbf a\|_\infty\leq1$. Therefore,
\begin{align}
d_{\mathcal F_1}
\bigl(
\rho(\mathbf r),\rho(\mathbf s)
\bigr)
&=
\sup_{\|\mathbf a\|_\infty\leq1}
\left|
(\mathbf r-\mathbf s)\cdot\mathbf a
\right|
\nonumber\\
&=
\|\mathbf r-\mathbf s\|_1.
\end{align}
Hence,
\begin{equation}
d_{\mathcal F_1}
\bigl(
\rho(\mathbf r),\rho(\mathbf s)
\bigr)
=
\|\mathbf r-\mathbf s\|_1.
\end{equation}

Finally, using $R(\rho) = \frac12 \operatorname{dist}_{\mathcal F}(\rho,\mathcal F)$, we obtain
\begin{equation}
R(\rho(\mathbf r))
=
\frac12
\inf_{\|\mathbf t\|_1\leq1}
\|\mathbf r-\mathbf t\|_1.
\end{equation}
The distance from $\mathbf r$ to the $\ell_1$-unit ball is
\begin{equation}
\inf_{\|\mathbf t\|_1\leq1}
\|\mathbf r-\mathbf t\|_1
=
\left(
\|\mathbf r\|_1-1
\right)_+.
\end{equation}
Therefore, $R(\rho(\mathbf r)) = \frac12 \left( \|\mathbf r\|_1-1 \right)_+$.

\subsection{Proof of the Continuous-Family Separation}
\label{app:single-qubit-family}

Consider $\mathbf r_A(a) = (a,a,0)$ and
\begin{equation}
\mathbf r_B(a)
=
\left(
\frac{2a}{3},
\frac{2a}{3},
\frac{2a}{3}
\right),
\end{equation}
with $\frac12<a<\frac1{\sqrt2}$.

\subsubsection{Mixedness and Equal RoM}

For $\rho_A(a)$, $\|\mathbf r_A(a)\|_2^2 = 2a^2 <1$, where the last inequality follows from $a<\frac1{\sqrt2}$. Thus, $\rho_A(a)$ is strictly mixed.
For $\rho_B(a)$,
\begin{equation}
\|\mathbf r_B(a)\|_2^2
=
3\left(\frac{2a}{3}\right)^2
=
\frac{4a^2}{3}
<1,
\end{equation}
throughout the same range. Hence, $\rho_B(a)$ is also strictly mixed.
Their $\ell_1$-norms coincide: $\|\mathbf r_A(a)\|_1 = 2a = \|\mathbf r_B(a)\|_1$. Since $a>\frac12$, both states lie outside the stabilizer octahedron. Using the single-qubit RoM
formula,
\begin{align}
R(\rho_A(a))
&=
\frac12(2a-1)
\nonumber\\
&=
a-\frac12,
\end{align}
and similarly $R(\rho_B(a)) = a-\frac12$. Therefore, $R(\rho_A(a)) = R(\rho_B(a)) = a-\frac12$.

\subsubsection{Forward Conversion}

Let $U$ be a single-qubit Clifford unitary satisfying $UXU^\dagger=Y, \qquad UYU^\dagger=Z, \qquad UZU^\dagger=X$, and define
\begin{equation}
\Lambda_{\mathrm{cyc}}(\rho)
=
\frac13
\left[
\rho
+
U\rho U^\dagger
+
U^2\rho(U^\dagger)^2
\right].
\end{equation}
Since $\Lambda_{\mathrm{cyc}}$ is a convex mixture of Clifford channels, $\Lambda_{\mathrm{cyc}}\in\mathrm{SPO}_1$.
On Bloch vectors, the channel acts as
\begin{equation}
(r_x,r_y,r_z)
\longmapsto
\left(
\frac{r_x+r_y+r_z}{3},
\frac{r_x+r_y+r_z}{3},
\frac{r_x+r_y+r_z}{3}
\right).
\end{equation}
For $\mathbf r_A(a)$, $r_{A,x}+r_{A,y}+r_{A,z} = 2a$. Hence, $\Lambda_{\mathrm{cyc}} \bigl( \rho_A(a) \bigr) = \rho_B(a)$. Therefore, $C_{\mathcal F_1} \bigl( \rho_A(a)\to\rho_B(a) \bigr) = 0$.

\subsubsection{Reverse Conversion: Primal Construction}

Define $\alpha := \frac{ \sqrt6+\sqrt2-\sqrt3-1 }{4}$ and $\beta := \frac{ 3+\sqrt3-\sqrt6-\sqrt2 }{4}$. These constants satisfy $\alpha+\beta=\frac12$.
Consider the affine map on Bloch vectors
\begin{equation}
\Lambda_\star:
(r_x,r_y,r_z)
\longmapsto
\left(
\beta+\alpha(r_x+r_y+r_z),
\,
\beta+\alpha(r_x+r_y+r_z),
\,
0
\right).
\end{equation}
We first verify that $\Lambda_\star\in\mathrm{SPO}_1$.
Let
\begin{equation}
\mathbf e_x=(1,0,0),
\qquad
\mathbf e_y=(0,1,0),
\qquad
\mathbf e_z=(0,0,1)
\end{equation}
denote the positive vertices of the stabilizer octahedron.
For every $i\in\{x,y,z\}$, $\mathbf e_i \longmapsto \left( \frac12,\frac12,0 \right)$, because $\alpha+\beta=\frac12$. For the negative vertices, $-\mathbf e_i \longmapsto (-\kappa,-\kappa,0)$, where $\kappa := \alpha-\beta = \frac{ \sqrt6+\sqrt2-\sqrt3-2 }{2}$. Moreover, $0<2\kappa<1$.
Therefore, $\left\| \left( \frac12,\frac12,0 \right) \right\|_1 = 1$ and $\|(-\kappa,-\kappa,0)\|_1 = 2\kappa <1$. Hence all six stabilizer vertices are mapped into $\mathcal F_1$. Since
$\mathcal F_1$ is their convex hull and $\Lambda_\star$ is affine, $\Lambda_\star(\mathcal F_1) \subseteq \mathcal F_1$.

It remains to verify complete positivity. Define $H:=X+Y, \qquad V:=X-Y+Z$. On the Pauli basis, $\Lambda_\star(I) = I+\beta H$, and $\Lambda_\star(X) = \Lambda_\star(Y) = \Lambda_\star(Z) = \alpha H$.
The Choi operator is therefore
\begin{equation}
J_{\Lambda_\star}
=
\frac12
\left[
I\otimes(I+\beta H)
+
\alpha V\otimes H
\right],
\end{equation}
or equivalently
\begin{equation}
J_{\Lambda_\star}
=
\frac12
\left[
I\otimes I
+
(\beta I+\alpha V)\otimes H
\right].
\end{equation}
The spectra of $V$ and $H$ are
\begin{equation}
\operatorname{spec}(V)
=
\{\pm\sqrt3\},
\qquad
\operatorname{spec}(H)
=
\{\pm\sqrt2\}.
\end{equation}
Furthermore, $\beta+\sqrt3\alpha = \frac1{\sqrt2}$. Hence the four eigenvalues of $J_{\Lambda_\star}$ are $0, \qquad 1$, and $\frac12 \left[ 1+\sqrt2 \left( \beta-\sqrt3\alpha \right) \right]$, $\frac12 \left[ 1-\sqrt2 \left( \beta-\sqrt3\alpha \right) \right]$, all of which are nonnegative. Thus, $J_{\Lambda_\star}\geq0$.
Moreover, since $\operatorname{Tr}H=0$, we have $\operatorname{Tr}_{\mathrm{out}}J_{\Lambda_\star} = I$. Therefore, $\Lambda_\star$ is CPTP. Combining this with stabilizer preservation, $\Lambda_\star\in\mathrm{SPO}_1$.
For the input $\rho_B(a)$, the sum of the Bloch coordinates is $2a$. Hence, $\Lambda_\star \bigl( \rho_B(a) \bigr) = \rho((u_a,u_a,0))$, where $u_a = \beta+2a\alpha$.
Using $\alpha+\beta=\frac12$, we obtain $a-u_a = (2a-1)\beta$. Therefore,
\begin{align}
d_{\mathcal F_1}
\left(
\Lambda_\star(\rho_B(a)),
\rho_A(a)
\right)
&=
2(a-u_a)
\nonumber\\
&=
2(2a-1)\beta.
\end{align}
It follows that $C_{\mathcal F_1} \bigl( \rho_B(a)\to\rho_A(a) \bigr) \leq (2a-1)\beta$.

\subsubsection{Reverse Conversion: Matching Dual Bound}

We next prove that the preceding upper bound is optimal.
Let $H=X+Y$. For an arbitrary output state $\rho(\mathbf s), \qquad \mathbf s=(s_x,s_y,s_z)$, the single-qubit metric gives
\begin{align}
d_{\mathcal F_1}
\bigl(
\rho(\mathbf s),\rho_A(a)
\bigr)
&=
|s_x-a|
+
|s_y-a|
+
|s_z|
\nonumber\\
&\geq
2a-(s_x+s_y)
\nonumber\\
&=
2a-\operatorname{Tr}[H\rho(\mathbf s)].
\end{align}
Therefore,
\begin{equation}
2C_{\mathcal F_1}
\bigl(
\rho_B(a)\to\rho_A(a)
\bigr)
\geq
2a-M_\star(a),
\end{equation}
where
\begin{equation}
M_\star(a)
:=
\sup_{\Lambda\in\mathrm{SPO}_1}
\operatorname{Tr}
\left[
H\Lambda(\rho_B(a))
\right].
\end{equation}
Since transposition changes the sign of $Y$, $\rho_B(a)^T = \frac12 \left( I+\frac{2a}{3}V \right), \qquad V=X-Y+Z$.
Define the three stabilizer states
\begin{equation}
\phi_x=\frac{I+X}{2},
\qquad
\phi_y=\frac{I+Y}{2},
\qquad
\phi_z=\frac{I+Z}{2},
\end{equation}
and $S := \phi_x^T+\phi_y^T+\phi_z^T = \frac{3I+V}{2}$.
For every $\Lambda\in\mathrm{SPO}_1$, we have $\Lambda(\phi_i)\in\mathcal F_1$. Since $\sup_{\tau\in\mathcal F_1} \operatorname{Tr}(H\tau) = 1$, it follows that
\begin{equation}
\operatorname{Tr}
\left[
H\Lambda(\phi_i)
\right]
\leq1,
\qquad
i\in\{x,y,z\}.
\end{equation}
Define $\mu_\star(a) := \frac{ 1-\frac{2\sqrt3a}{3} }{ 3-\sqrt3 }$ and $A_\star(a) := \mu_\star(a)S-\rho_B(a)^T$. The eigenvalues of $A_\star(a)$ are $0$ and $-\frac{1+\sqrt3}{2}(2a-1)$. Since $a>\frac12$, we have $A_\star(a)\leq0$.
Define $Q_\star(a) := -\sqrt2A_\star(a) \geq0$. The dual slack operator
\begin{equation}
D_\star(a)
:=
Q_\star(a)\otimes I
+
\mu_\star(a)S\otimes H
-
\rho_B(a)^T\otimes H
\end{equation}
factorizes as $D_\star(a) = [-A_\star(a)] \otimes (\sqrt2I-H) \geq0$.
For the Choi operator $J_\Lambda$ of any
$\Lambda\in\mathrm{SPO}_1$, we use
\begin{equation}
\operatorname{Tr}
\left[
(\eta^T\otimes W)J_\Lambda
\right]
=
\operatorname{Tr}
\left[
W\Lambda(\eta)
\right].
\end{equation}
Together with $D_\star(a)\geq0,~ J_\Lambda\geq0$, the trace-preserving condition, and the stabilizer constraints above, this gives
\begin{equation}
\operatorname{Tr}
\left[
H\Lambda(\rho_B(a))
\right]
\leq
\operatorname{Tr}[Q_\star(a)]
+
3\mu_\star(a).
\end{equation}
A direct simplification yields $\operatorname{Tr}[Q_\star(a)] + 3\mu_\star(a) = 2a-2(2a-1)\beta$. Moreover, the primal channel $\Lambda_\star$ attains this value because
\begin{align}
\operatorname{Tr}
\left[
H\Lambda_\star(\rho_B(a))
\right]
&=
2u_a
\nonumber=
2\beta+4a\alpha
\nonumber=
2a-2(2a-1)\beta.
\end{align}
Hence, $M_\star(a) = 2a-2(2a-1)\beta$.
Substituting this into the distance bound gives
\begin{align}
2C_{\mathcal F_1}
\bigl(
\rho_B(a)\to\rho_A(a)
\bigr)
&\geq
2a-M_\star(a)=
2(2a-1)\beta.
\end{align}
Therefore, $C_{\mathcal F_1} \bigl( \rho_B(a)\to\rho_A(a) \bigr) \geq (2a-1)\beta$.
This matches the primal upper bound, and hence $C_{\mathcal F_1} \bigl( \rho_B(a)\to\rho_A(a) \bigr) = (2a-1)\beta$. Substituting $\beta = \frac{ 3+\sqrt3-\sqrt6-\sqrt2 }{4}$ finally gives
\begin{equation}
C_{\mathcal F_1}
\bigl(
\rho_B(a)\to\rho_A(a)
\bigr)
=
(2a-1)
\frac{
3+\sqrt3-\sqrt6-\sqrt2
}{4}
>0.
\end{equation}
For $a=\frac35$, this reduces to
\begin{equation}
C_{\mathcal F_1}
(\rho_B\to\rho_A)
=
\frac{
3+\sqrt3-\sqrt6-\sqrt2
}{
20
}.
\end{equation}

\bibliography{ref} 
\bibliographystyle{unsrt} 

\end{document}